\documentclass[twocolumn, amsmath,amssymb, superscriptaddress]{revtex4}

\usepackage{dcolumn}
\usepackage{bm}
\usepackage{amsmath}
\usepackage{amsfonts}
\usepackage{graphics} 
\usepackage[dvips]{graphicx}
\usepackage{times}
\usepackage{setspace}
\usepackage{color}   
\usepackage{verbatim}
\usepackage[raggedright]{titlesec}
\usepackage[normalem]{ulem}
\usepackage{framed}

\usepackage{enumerate}
\usepackage[breaklinks=true]{hyperref} 
\usepackage{breakurl} 

\usepackage{balance}

\hypersetup{colorlinks=true,citecolor=blue, linkcolor=black,urlcolor=blue} 

\begin{document}

\title{The Memory of Science: Inflation, Myopia, and the Knowledge Network}

\author{Raj K. Pan$^{a,}$}
\affiliation{Department of Computer Science, Aalto University School of Science, P.O. Box 15400, FI-00076, Finland}
\author{Alexander M. Petersen$^{a,b,}$}
\affiliation{Laboratory for the Analysis of Complex Economic Systems, Institutions Markets Technologies (IMT)  School for Advanced Studies Lucca, Lucca 55100, Italy}
\affiliation{Management Program, School of Engineering, University of California, Merced, California 95343}
\author{Fabio Pammolli}
\affiliation{Laboratory for the Analysis of Complex Economic Systems, Institutions Markets Technologies (IMT)  School for Advanced Studies Lucca, Lucca 55100, Italy}
\author{Santo Fortunato$^{b,}$}
\affiliation{Center for Complex Networks and Systems Research, School of Informatics and Computing, Indiana University, Bloomington, IN, USA}

\begin{abstract}
Science is a growing system, exhibiting  ~4\% annual growth in publications and  ~1.8\% annual growth in the number of references per publication. Together these growth factors correspond to a 12-year doubling period in the total supply of references, thereby challenging traditional methods of evaluating  scientific production, from researchers to institutions. 
Against this background, we analyzed a citation network comprised of  837 million references produced by 32.6 million  publications over the period 1965-2012, allowing for a detailed analysis of the `attention economy'  in science. Unlike previous studies, we analyzed the entire probability  distribution of reference ages -- the time difference between a citing and cited paper -- thereby capturing previously overlooked trends. Over this half-century period we observe  a narrowing range of attention --  both classic and recent literature are being cited increasingly less, pointing to the important role of socio-technical processes. 
To better understand how these patterns fit together, we developed a network-based model of the scientific enterprise, featuring exponential growth, the redirection of scientific attention via publications' reference lists, and the crowding out of old literature by the new. We validate the model against several empirical benchmarks. We then use the model to test the causal impact of paradigm shifts in science, thereby providing theoretical guidance for science policy analysis. 
In particular, we show how  perturbations to the growth rate of scientific output -- i.e. following from the new layer of rapid online publications --  affects the reference age distribution and the functionality of the vast science citation network as an aid for the search \& retrieval of knowledge. In order to account for the inflation of science, our study points to the need for a systemic overhaul of the counting methods used to evaluate citation impact -- especially in the case of evaluating science careers, which can span several decades and thus several doubling periods.
\end{abstract} 



\maketitle

\footnotetext[1]{ These authors contributed equally.}
\footnotetext[2]{Send correspondence to santo.fortunato@gmail.com\\  or petersen.xander@gmail.com}

Driven by  public and private sector investment
into people and projects \citep{stokes_pasteurs_1997,stephan_how_2012},   the rate of scientific production has exhibited persistent growth over the last century \citep{lariviere_long-term_2008,ASI:ASI20936}. 
However, 
the existing literature seems to neglect, or at least underestimate, the impact of long term growth trends on measurements made across different historical periods. As recently demonstrated in a study of attention decay in citation life cycles, controlling for growth rates can drastically change  measured trends
 \citep{parolo_attention_2015}.
 This example highlights the need for a better understanding of how the scientific attention economy \citep{Franck_Scientific_1999,Klamer_Attention_2002} is impacted by growth of the scientific system. Moreover, the recent proliferation of a new layer of rapid-publication 
 ``e-journals'' has contributed considerably to this growth, making it a  pressing and relevant issue. 
 
Large historical databases of research output provide scientists the opportunity to study themselves in an emerging research domain identified as the  `science of science', which aims to provide valuable insights for science policy \citep{fealing_science_2011} in addition to the science of complex systems.
Indeed, millions of publications are now produced each year by scientists around the world, providing   quantifiable  links to the past. These links are preserved  within each publication as bibliographic references, which  provide  a means to measure how much today's research builds upon yesteryear's.
As such, the  citation network  --  where  nodes are publications and links are the references within a publication to prior literature --  has  been used to conceptualize and measure the processes underlying the  evolution of the scientific enterprise for more than half a century \citep{garfield_citation_1955,deSollaPrice_networks_1965},  and continues to be useful for making important insights into the long-term evolution  of the scientific enterprise \citep{100yearsphysics}. 

Against this background we analyze the interplay between output growth and the memory in this complex system, simultaneously identifying and modeling three key mechanisms of the citation network: 
\begin{enumerate}
\item the exponential growth of  the  total number of references produced each year due to the growth in both publication output and reference list length (inflation), 
\item the subsequent concentration of citations received by publications (citation, or attention, inequality), and
\item the distribution of references backwards in time (obsolescence of  knowledge).
\end{enumerate}
These considerations are important in the measurement, interpretation, and modeling of science for three fundamental reasons.
First, inflation in the supply of references  affects any counting scheme of citations received, thereby impacting the comparative evaluation of careers, institutions, and country output across different periods.  While the bibliometrics community has certainly considered normalization strategies for comparing citation measures between varying time periods and disciplines, the existing strategies do not consider what we identify as the  ``inflation problem'' that is encountered when combining citation counts over time  into arbitrary totals (i.e. longitudinally  aggregated measures).  In order to remove the underlying bias arising from the steady inflation in scientific output (i.e. secular trends), we show that temporal discount factors are necessary. We demonstrate our normalization method on a large set of research careers from a wide range of age cohorts.

Second, the increasing supply of references has dramatically diminished the fraction of publications that go uncited. While this shift may at first appear to be  an incremental change in the lower tail of the citation distribution, it has an enormous impact on the overall connectivity of the citation network, thereby affecting the search and retrieval of knowledge.

Third, there is the question of whether, and to what extent, a diminishing depth and breadth of attention  is related to the system's growth, and whether a  decreasing attention  to older literature may negatively impact the efficiency of the knowledge production function.\footnotetext[3]{Nevertheless, information and communications technology (ICT) innovations   may compensate this imperfection  by improving researchers' ability to search, retrieve, store, and recall knowledge  \citep{Sparrow05082011}.}

\vspace{-0.2in}
\section*{Literature review and   framework}
\vspace{-0.2in}
\noindent By considering the three features simultaneously -- each of which has been addressed separately in the literature, however with some notable disagreement -- we aim to show how they are all related to the inflation of science.
First, however, we shall outline the  fragmentation of the existing  literature pertaining to  these three features. 

While the general  consensus is that the rate of uncited publications is declining   \citep{schwartz_rise_1997,Wallace2009296,lariviere_decline_2009},
the level of inequality in the citation distribution has been shown to either decline   \citep{acharya_rise_2014, petersen_inequality_2014,lariviere_decline_2009} or increase \citep{Barabasi2012,Evans18072008}, depending  on the method and the perspective.
There is also disagreement concerning the  obsolescence rate of scientific literature -- used as a quantitative proxy to estimate the life-cycle of knowledge.
For example,   \cite{Evans18072008} finds that journals with more availability of online back-issues tend to have reference lists that are more focussed on recent literature (more myopic), explained as the result of the  availability of efficient online hyperlinks that mediate the browsing  of the publications listed in reference lists.
Meanwhile, two recent studies  report that older publications  are being cited (as a percent) more and more over time 
 \citep{wallace_small_2012,verstak_shoulders_2014}. 
These discrepancies   demonstrate the need for a methodological framework that accounts for the systematic bias introduced by the exponential increase of scientific output. Indeed, although  the inflation of scientific output has been a long-standing issue  \cite{deSollaPrice_networks_1965,Broad_publishinggame_1981}, only recently have analogs of  inflation indices been used to normalize impact factors \citep{ASI:ASI20936} and individual  citation counts \citep{petersen_reputation_2014}.
  
\begin{figure*}
\centering{\includegraphics[width=0.5\textwidth]{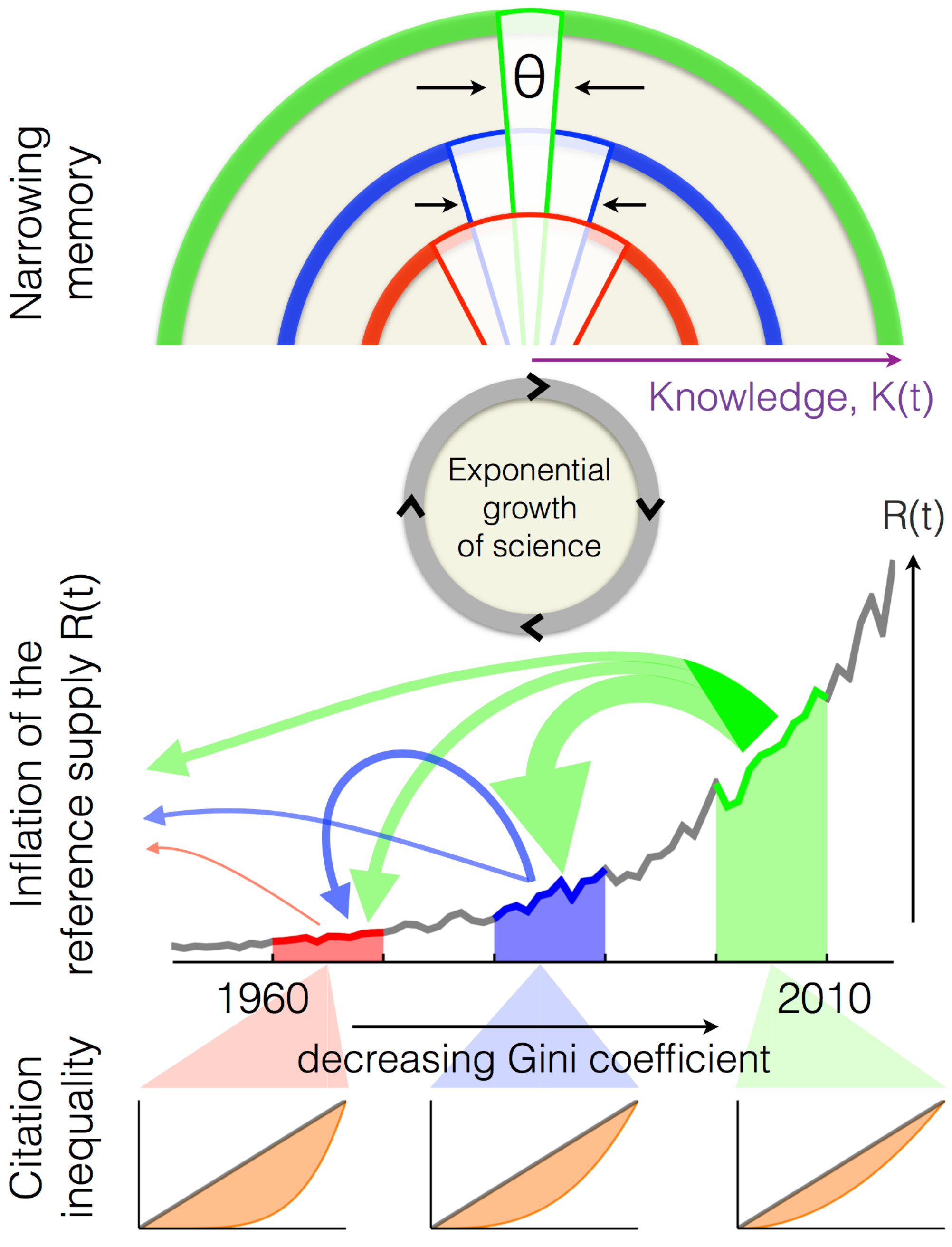}}
\caption{\label{scienceschem}  {\bf Complementary perspectives on the growth of the scientific attention economy.} The exponential increase in publications and reference list length means that more citations are made today than in the past -- also on a per-publication basis. The  citation of prior literature  is also growing nonlinearly with time, corresponding to  a retroactive intergenerational effect (i.e. the variable citation flows represented by the arrows in the middle panel). This inflation of the system impacts not only the concentration of citations received (bottom panel), but it also  is related to the narrowing of the range of expertise ($\theta$)  at the knowledge frontier as the  knowledge radius $K(t)$ increases (top panel). }
\end{figure*}

To address these issues we analyzed the Thompson Reuters Web of Knowledge (TR) publication index from 1965 to 2012, comprising $32.6$ million  publications and $838$ million references made (or from the alternative perspective, citations received).\footnotetext[4]{In our citation network analysis we consider  the obsolescence problem from both the  prospective (forward looking or diachronous) 
 as well as the retrospective (backward looking or synchronous) 
 perspectives   \citep{Nakamoto_synchronous_1988,Glanzel_forwardbackward_2004}.} We control for  disciplinary variation by  grouping the publication data using the three major subcategories defined by TR: (Natural) Science,  Social Sciences, and Arts \& Humanities (A\&H).  For each subject area, we analyzed the impact of the exponentially growing system (inflation) on the concentration of citations within the citation network (inequality) and the subsequent impact on the obsolescence of knowledge (narrowing memory), together illustrated in  Fig. \ref{scienceschem}.

We start by analyzing how the inflation of the supply of references made by individual publications has  affected the distribution of citations received across different publication year cohorts. For example, we measured a 5.6\% growth rate in $R(t)$, the total number of references produced (see Fig.  \ref{GrowthRtEtPtRtCt}), meaning that the total number of citations in the TR citation network doubles roughly every 12 years!
These basic considerations then lead naturally to the question: is the concentration of citation received increasing or decreasing? We find that the answer to  this question is intrinsically linked to the decreasing share of uncited publications, which is inherently related to the increasing supply of references produced.

 We then move to the question of whether the statistical patterns of citing new and old literature are changing over time. To this end, we analyze the temporal distances between a referencing publication and the cited publication, denoted as the reference distance  $\Delta_{r}$, by focusing on the shifts in the entire probability distribution $P(\Delta_{r})$  across 10-year intervals from 1970-2010.

Among our main findings is showing that the share of references going to the middle range of the  $P(\Delta_{r})$ distribution -- towards literature older than $\approx$ 6 y ($\Delta_{r}^{-}$) and less than  $\approx$ 50 y old ($\Delta_{r}^{+}$) -- has increased on average by 25\% over the study period for the three subject areas. 
Considered from an alternative perspective,  we observe a stable ``fixed point'' in the  lower end of $P(\Delta_{r})$, around $\Delta^{-}_{r}\approx 6$ years, which serves as  a useful benchmark for classifying academic literature as contemporary ($<6$ y) or established ($\geq 6$ y).   

Interestingly, the decline in the percentage of references for $\Delta_{r} < \Delta_{r}^{-}$ may follow from the rapid expansion of the system, whereby the trade-off between short-term and long-term memory are stressing cognitive limits.
There may also be social cohort effects within scientific communities, wherein the incentives to follow established leaders may reinforce the concentration of attention away from very recent as well as very distant research. Indeed, scientific reputation is a social mechanism that serves  as an `identification device' to aid with the  information  overload problem, and its role may be becoming stronger as quantitative measures become increasingly  prevalent in science \citep{petersen_reputation_2014}.

To better understand these and other empirical trends, we develop a network-based citation model which incorporates four  key ingredients that capture both  real features of the academic citation system and the process that authors follow while searching within and traversing across the knowledge network: 
\begin{enumerate}[(i)]
\item exponential growth in the number of publications published each year, $n(t)$, and the number of references per publication, $r(t)$, 
\item  crowding out of old literature by new literature, which we impose using an attention bias operationalized by  $n(t)$ itself, 
\item a   citation mechanism (link-dynamics) capturing the  orientation  of scientific attention towards high-impact literature (a positive feedback mechanism),  and 
\item a redirection link-formation process that is inspired by the now common behavior of  finding related knowledge by following  the reference list of a source article. 
\end{enumerate}
Because (iv) is implemented via a tunable  parameter controlling the rate of  triadic closure in the citation network, this model falls into the class of  redirection models \citep{Holme_Growing_2002,vazquez_growing_2003,krapivsky_organization_2001,krapivsky_network_2005,gabel_highly_2013,gabel_highly_2014}. 
We show that our generative model reproduces various stylized facts observed for the empirical citation network (elaborated further in the online-only {\it Supplementary Material (SM)} text).
Moreover, our  model provides the opportunity to accurately explore the  causal impact of several scenarios faced by science, and thus, to provide  informative  insights for science policy makers concerned with the impact of growth trends on the evolution of science.

 We close our results with an empirical `real world' demonstration by analyzing the publication profiles of 551 researchers, showing that ignoring inflation when calculating two common cumulative measures -- a researcher's total citations and his/her $h$-index --  can lead to a significant underestimation in the quantitative evaluation of scientific achievement.  To be specific, we demonstrate this problem  by using an intuitive deflation index to normalize productivity and impact measures of individual scientists from different age cohorts, and compare the raw measures to those which have been ``discounted''. We conclude with a discussion of our results and some policy implications.

\vspace{-0.2in}
\section*{Material and Methods}
\vspace{-0.2in}
\noindent We analyzed all English publications (articles and reviews) between 1965 through 2012 in the Thomson Reuters Web of Knowledge (TR) database. We  used the TR  journal classification  to separate the publication data  into 3 broad domains: (natural) ``Science'', ``Social Sciences'', and ``Art \& Humanities''. We focus our analysis on the Science and Social Science domains, which account for more than 95\% of the data. We leave the possibility of analyzing patterns within individual ``subject areas'' (as defined by TR) as an open avenue of potential future research.
Further details on the data can be found in the  SM text.


For a given publication $p$ published in year $t$, we collected the set of references made by $p$. Then, for a given reference $r$ with  publication year $t_{r}$, we define the reference distance as  $\Delta_{r}\equiv t-t_{r}$, i.e. the distance in time between $p$ and $r$.
 Using the references we also calculated the total citation count 
$c^{p}$ for each $p$.

Country-level R\&D indicator data was obtained from the World Bank \citep{WorldBankData}. 
The 551 researchers and their publications were identified and disambiguated using the TR ``distinct author`` database feature combined with TR {\it ResearcherID.com} profiles.
For further details on the selection of the researchers analyzed here, see the {\it SM} Text.
 
\vspace{-0.2in}
\section*{Results} 
\begin{figure*}
\centering{\includegraphics[width=0.99\textwidth]{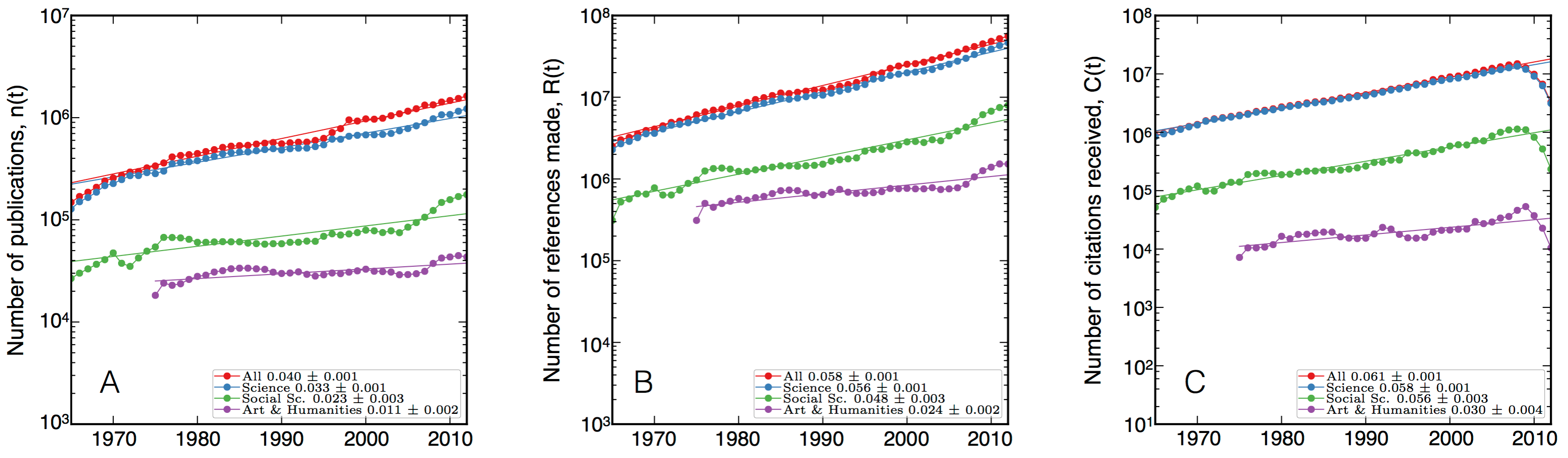}}
\caption{\label{GrowthRtEtPtRtCt}  {\bf Outputs of  scientific R\&D: empirical growth trends.} 
 Empirical growth trends of scientific output: publications and references. (A) Growth in the number of publications per year, $n(t)$. (B) Growth in the total supply of (outgoing) references per year, $R(t)$. (C) Growth in the total number of incoming citations per year, $C(t)$ (as measured in our citation census year $Y=2012$). Interestingly,  $C(t)$ is growing faster than $R(t)$, indicating that more references are being concentrated on publications from more recent years.}
\end{figure*}

\begin{figure*}[h]
\centering{\includegraphics[width=0.83\textwidth]{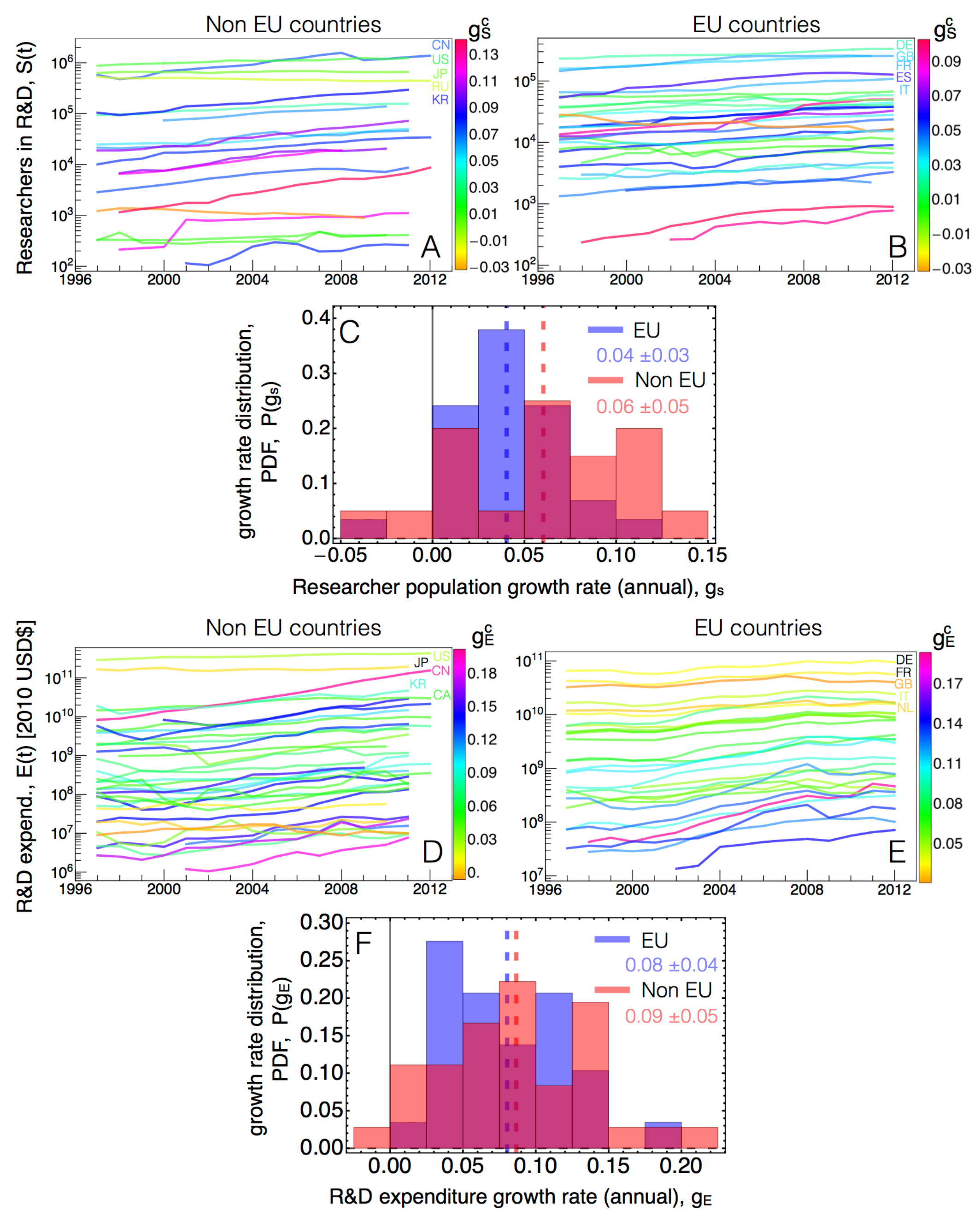}}
\caption{\label{RnDInputGrowth} {\bf Inputs  of  scientific R\&D: empirical growth trends.}  (A-F) Empirical growth trends in the researcher population and funding over the 16-year period  1997-2012. (A-C) Growth of the number of researchers in R\&D, $S^{c}(t)$,  by country $c$. (D-F) Growth in the R\&D expenditure, $E^{c}(t)$,  by country $c$. Only countries with more than 10 data points are analyzed. (A) Time series' of $S^{c}(t)$ for 20 large non European Union countries.   (B) Time series' of $S^{c}(t)$ for 29 European Union and EFTA countries.  (C) Frequency distribution of the exponential growth rate $g^{c}_{S}$, estimated for each $S^{c}(t)$ time series using ordinary least squares regression. The mean (dashed vertical line) and std. dev. for each country subgroup  are shown in the panel.
(D) Time series' of $E^{c}(t)$ for 36 large non European Union countries.  (E) Time series' of $E^{c}(t)$ for 29 European Union and EFTA countries.  (F) Frequency distribution of the exponential growth rate $g^{c}_{E}$, estimated for each $E^{c}(t)$ time series using ordinary least squares regression. 
We tested the difference in mean (Student T-test), median (Mann-Whitney test), and distribution (Kolmogorov-Smirnov test)  between the EU and non-EU growth rates, finding there to be no statistically significant difference ($p>0.59$ in each test).
(A,B,D,E) Each color legend indicates the growth value corresponding to each individual time series.
 For comparison, growth rate of   the population size of post doctorates and graduate students in U.S. STEM fields is also growing roughly 2-4\% over the time period 1972-2010  \citep{petersen_quantitative_2014}. 
Country level  data calculated from World Bank sources \citep{WorldBankData}: \href{http://data.worldbank.org/indicator/SP.POP.SCIE.RD.P6/countries}{Researchers in R\&D (per million people)} and \href{http://data.worldbank.org/indicator/GB.XPD.RSDV.GD.ZS}{Research and development expenditure (\% of GDP)} data combined with \href{http://data.worldbank.org/indicator/NY.GDP.MKTP.CD/countries}{GDP (current US\$)} and \href{http://data.worldbank.org/indicator/SP.POP.TOTL/countries}{Total Population (in number of people)} data. All dollar amounts deflated to 2010 US\$.}
\end{figure*}

\vspace{-0.2in}
\noindent{\bf Scientific growth and the inflation of the number of citations.} 
Many leading countries fund  R\&D activities as a percentage of their GDP, thereby sustaining the growth and diversification of scientific effort  \citep{petersen_quantitative_2014}.
By way of example,  Fig. \ref{RnDInputGrowth} shows the growth of the principal inputs of the scientific enterprise -- people and money -- from 1997-2012. Because the EU has its own collective funding system, we separated the countries into two groups, Non-EU and EU. Interestingly, when considering the sustained growth of the inputs, specifically  R\&D labor and government expenditure on R\&D, it becomes rather clear that China has emerged as a global leader in the production of scientific knowledge \citep{Zhou200683} by investing in the growth of its researcher population.
Furthermore, Figs. \ref{GrowthRtEtPtRtCt} and \ref{RnDInputGrowth} 
 indicate that the supply of references  and the growth of  the R\&D researcher population  are  growing at roughly the same annual rate of 5-6\%.

The unit of analysis in our study  are publication-to-publication associations --  termed ``references'' from the  outgoing perspective (i.e. reference list) and ``citations'' from the incoming perspective (i.e. citation count $c^{p}$ of a given publication $p$). We then measure the growth of the reference supply drawing from two sources: (i) the increasing number of publications  $n(t)$ produced per year $t$,  and (ii) the increasing (average) number of references per publication, $r(t)$. 
For example, for the publications belonging to Science,   we measure an annual exponential growth rate $g_{n}=0.033$ for publications and $g_{r}=0.018$ for the   references per publication. As a result, the net annual reference supply, $R(t)$, is also increasing exponentially with annual growth rate $g_{R} \approx g_{n}+g_{r}$. For example, using the time series for the total references produced in a given year, $R(t)$, we measured $g_{R}=0.056 \pm 0.001$  for Science (see Fig. \ref{GrowthRtEtPtRtCt}). 
\footnotetext[5]{This increase in the citation credit supply is analogous to monetary inflation in economics.  Interestingly, De Solla Price estimated a publication doubling time of 13.5 years, corresponding  $g_{n} = \ln(2)/13.5 y = 0.05 y^{-1}$ in 1965 using publication data  for 1862-1961 \citep{deSollaPrice_networks_1965}. For comparison, here we use $R(t)$ to estimate the rate of growth of the number of connections in the knowledge network, finding the  doubling period  to be roughly $\ln(2)/0.056 \approx 12$ y.}


One main difficulty in   analyzing citation statistics  is  the fact that references produced later in time can impact the citation tallies of publications earlier in time -- a backwards inter-generational flow of the reference supply. Because the doubling period for the reference supply is only 12 years, it is thus conceivable that some publications have an extended citation lifecycle across several decades merely due to the underlying  inflation.
Thus, in order to control for this retroactive effect, we tallied the citation counts $c^{p}_{t, \Delta t}$ for each publication $p$ from year $t$ using only the citations arriving in the fixed 5-year ($\Delta t= 5$) window $[t, t+\Delta t]$ (note that we  only use $c^{p}_{t,\Delta t=5}$  in this and the next subsection). In this way, we limit the impact of the long-term increase in the supply of references across time.

\begin{figure*}
\centering{\includegraphics[width=0.7\textwidth]{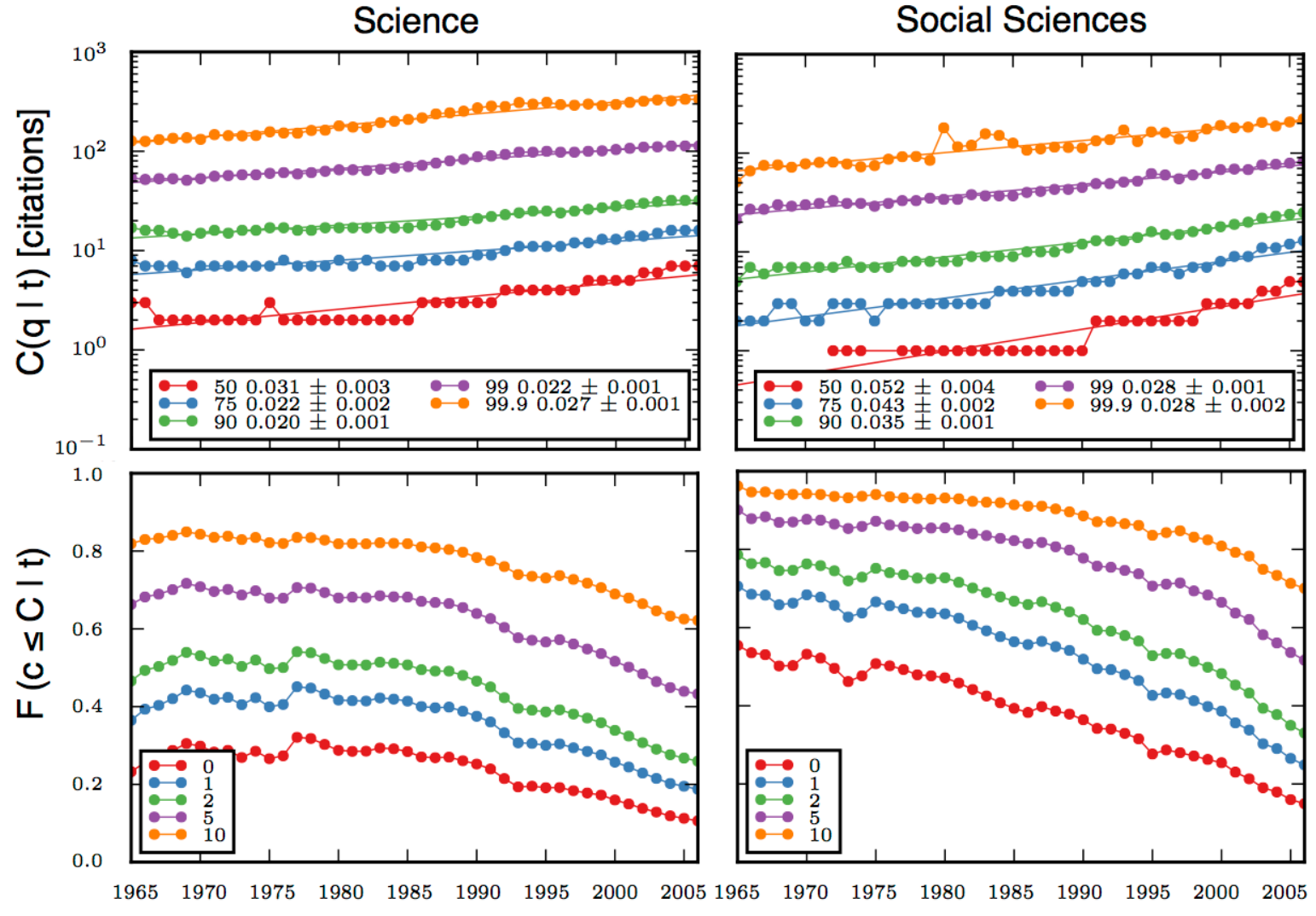}}
\caption{\label{CqtUnctFdrt}   {\bf Empirical  trends in the upper and lower  tails of the citation distribution.} (top row) Increase in the broadness of the citation distribution. The citation value $C(q \vert t)$ corresponding to a given percentile $q$ of the citation distribution $P(c^{p}_{t,5})$. Each line of the legend shows two numbers: the percentile value  100$\times q$ and the best-fit exponential growth parameter calculated for each curve.  (bottom row) Decrease in the fraction of lowly-cited publications. Each curve represents the fraction of publications with $c^{p}_{t, 5}\leq C$ citations received,  for each threshold $C$ and for each year.  
For Arts \& Humanities see Figs. S2-S5.}
\end{figure*}

We then calculated   $C(q \vert t)$, the citation value corresponding to a given percentile $q$ of the citation distribution, for each year $t$.  For example,  Fig. \ref{CqtUnctFdrt} shows that the top 1\% of publications from $t=$2000 had more than  100 citations as of 2005, whereas the top 1\% of publications from 1965 had only 50 citations as of 1970.  Each $C(q \vert t)$ is growing at a slow exponential rate $\lesssim g_{n}$,  which is larger for smaller $q$, pointing to a decreasing concentration of citations which we will address in the next section. Because we only use citations within the 5-year window, these trends are not sensitive to  long-term trends in the obsolescence rate of scientific publications \cite{parolo_attention_2015}. 
In all, the steady growth of $C(q \vert t)$  illustrates how the {\it relative} ``value'' of citations is systematically decreasing over time, i.e. the entire citation distribution is systematically shifting towards higher values. In other words, more recent publications need increasingly more  citations to be within the top 5\% (of publications from the same year)  than do older publications  in order to achieve the same percentile value within their respective  publication cohort.\\ 

\noindent{\bf Impact of growth on uncitedness and citation inequality.} 
We calculated the fraction $F(c\leq C \vert t)$ of publications with $c^{p}_{t, 5}\leq C$ citations received, for the range $0 \leq C \leq 10$ (e.g. the threshold $C=0$ corresponds to uncited publications). Figure \ref{CqtUnctFdrt} shows that the $F(c\leq C \vert t)$ are all decreasing, pointing to the relation between sustained growth of the references supply and decreasing citation inequality. 
For example, in 1980, roughly 30\% of Science publications remained uncited 5 years after publication.
By 2005, this percentage decreased to roughly 10\%. Moreover, roughly 60\% of Science publications  from 2005 have 10 or less citations after 5 years.
This decreasing trend has occurred in Science since the 1980s and in Social Science since at least the mid 1960s.
Note that the gap between the curves for $C=10,5,2$ is approximately constant over the last 20-30 years, indicating that the share of references {\it within} these ranges is not changing dramatically. Thus,  the largest decrease over time is for the fraction of publications with $C=0,1,2$ citations, in that order. 



Subtle changes in the fraction of  uncited publications can have a  dramatic effects on the connectivity of the citation (knowledge) network. In practice, this topological alteration can have further  impact the functionality of online search \& retrieval algorithms, such as {\it Google Inc.'s} PageRank method \citep{PageRank:1998},  which is based on random walkers traversing the underlying information network, and is a common method implemented in citation network studies. 

Moreover, subtle shifts in the rate of uncited publications  can have pronounced effect on the concentration of citations in the citation distribution. 
In order to demonstrate this relation, we calculated the  Gini inequality coefficient 
\begin{equation} G(t) =  (2 n^{2}(t) \langle c(t) \rangle)^{-1} \sum_{i=1}^{n(t)}  \sum_{j=1}^{n(t)} \vert c_{t}^{i}- c_{t}^{j} \vert \ , 
\end{equation}
calculated for each year $t$, where $i$ and $j$ are indices running over the  set of $n(t)$ papers from each publication  cohort $t$ and $\langle c(t) \rangle$ is the mean number of citations among $p$ from $t$ \citep{Dixon_Gini_1987}. $G(t)$ is a relative (intensive) measure of the  pairwise difference between all data values in the sample, and as such,  outperforms other extensive inequality measures, such as the Herfindahl-Hirschman Index  $HHI(t)=n(t) \langle c^{2}_{t}\rangle/C^{2}(t)$.  
\footnotetext[6]{The HHI index is the product of the sample size, $n(t)$, the second moment of the distribution, $\langle c^{2}_{t}\rangle$, divided by  the net number of citations received by $p$ from $t$, $C(t)$, squared. Thus, unlike $G(t)$ which is an intensive measure of the  mean differences, the HHI is extensive (its value depends on the system size) as it confounds sample size bias due to the growing system with sensitivity to extreme values which are typical of citation distributions.
}
 $G(t)$ is less sensitive to fluctuations and to system size bias because it incorporates information from the entire citation distribution (all moments instead of just the second moment, $\langle c^{2}_{t}\rangle$).
Moreover,  it is a standardized distribution measure  with values ranging from 0 ($c^{p}_{t,5} =$ const. $\forall p$) to 1 (extreme inequality, i.e. all publications  have $c^{p}_{t,5} = 0$ except for one). 

Our analysis reveals a slow but substantial decrease in $G(t)$ (see Fig. S4).
 To test whether the decreasing $G(t)$ is related to the decreasing trend in $F(c= 0 \vert t)$, we recalculated $G(t)$ ignoring the uncited publications, finding the  negative trend in $G(t)$ to be subsequently reduced. Moreover, the level of inequality was also reduced when ignoring uncited publications in the calculation of $G(t)$. 
 
 Thus, the vanishing of uncitedness, induced by the increasing supply of references,  is sufficient to explain why  citation  inequality has decreased.
 To ensure that this trend  is not just due to the expanding coverage  of  TR -- which could    artificially decrease $G(t)$ due to a statistical bias arising from the indexing of more journals with on average lower impact factors --  we also verified the decreasing trend in  $G(t)$ using only $p$ from the high-impact journals {\it Nature}, {\it PNAS}, and {\it Science}  (see also Fig. S4).\\

\noindent{\bf Quantifying  the obsolescence of scientific literature.} 
We searched for quantitative trends in the citation lifecycle by analyzing  a large set of 837,596,576 references, defining the reference distance ($\Delta_{r}\geq 0$) as the time difference between the  publication date of the referencing publication  and the cited publication. We note that from this section forward, we include all citations, even those received after $\Delta_{t} = 5$ y, thus breaking from the method of the previous two sections which only considered citations received through the first 5 years. 

Over the period 1965-2012 we observe a clear decline in the fraction $F(\Delta_{r} \leq 3 \vert t)$ of  references less than  3 years old, and a small yet significant increase in $F(\Delta_{r} \leq 50 \vert t)$ for Science and Social Sciences and $F(\Delta_{r} \leq 30 \vert t)$  for A\&H  (See Fig. S5). The trends are subtle but significant in their net impact, the former implying a decreased attention to very recent literature and the latter implying a decreased attention to very old literature. 

To obtain a more comprehensive understanding, we analyzed the entire probability distribution $P(\Delta_{r}  \vert t)$ and cumulative probability distribution $CDF(\geq \Delta_{r}  \vert t)$, shown in Fig. \ref{DistDr}A,  at  10 year intervals over the period 1970--2010. 
 Interestingly, we find that the  $P(\Delta_{r}  \vert t)$ within each discipline intersect around the value $\Delta^{-}_{r} \approx 6$ years, despite the growth in $n(t)$ and $r(t)$. 
We also observe a second crossing point  in  $CDF(\geq \Delta_{r}  \vert t)$ around $\Delta^{+}_{r}\approx 50$ (Science), 20 (Soc. Sciences), and 40 years (A\&H). Thus, by analyzing the entire range of $\Delta_{r}$ it is evident that the decreasing $F(\Delta_{r} <\Delta^{-}_{r})$ and  $F(\Delta_{r}>\Delta^{+}_{r})$ is complemented by the redistribution of $\Delta_{r} $ into the reference distance interval $[\Delta^{-}_{r},\Delta^{+}_{r}]$ y.
In the following Model section we use the simulation of a theoretical network model to better understand the decreasing trend in extreme myopia and  extreme hyperopia, and in the Discussion section we discuss hypothetical behavioral and technological  explanations for these trends.\\
 
\noindent{\bf Network growth model featuring inflation, obsolescence, and myopia.}  
Stochastic growth models can provide mechanistic insights into the evolution of competition and growth in various complex systems  \citep{Buldyrev_growth_2007,petersen_quantitative_2011,petersen_persistence_2012,Scharnhorst_models_2012,Golosovsky_transition_2013,Golosovsky_uncovering_2014,petersen_reputation_2014}. 
In order to gain mechanistic insights into the impact of growth on citation inequality, uncitedness, and  shifts in  $P(\Delta_{r}  \vert t)$, we developed a generative science citation network model which we implement using Monte Carlo simulation. 

The initial conditions of our synthetic science  system is a small batch of $n(0)\equiv10$ nodes (publications), each with no outgoing references.
Then, in each time step $t=1...T$ a cohort of $n(t)$ nodes (indexed by $j$) are added to the pre-existing system of nodes. We matched the slow exponential growth of $n(t)$ and $r(t)$ to empirical data, using $g_{n}^{model} \equiv 0.033$ and $g_{r}^{model}\equiv 0.018$, so that $g_{R}^{model} =  g_{n}^{model} + g_{r}^{model} =  0.051$, thereby using $g_{n}$ and $g_{r}$ as the fundamental growth parameters for the simulation.

Our model is also designed to capture the behavioral aspects of synthesizing reference lists, wherein article browsing is facilitated by using the reference lists themselves as additional pathways to browse and find other articles. 
As such, this  network growth model   falls into the class of redirection models 
\citep{Holme_Growing_2002,vazquez_growing_2003,krapivsky_organization_2001,krapivsky_network_2005,gabel_highly_2013,gabel_highly_2014}, but is distinguished from other models  by its relatively high rate of triadic closure \citep{Holme_Growing_2002,PhysRevE.80.037101,ren_modeling_2012}. 

\begin{figure*}
\centering{\includegraphics[width=0.99\textwidth]{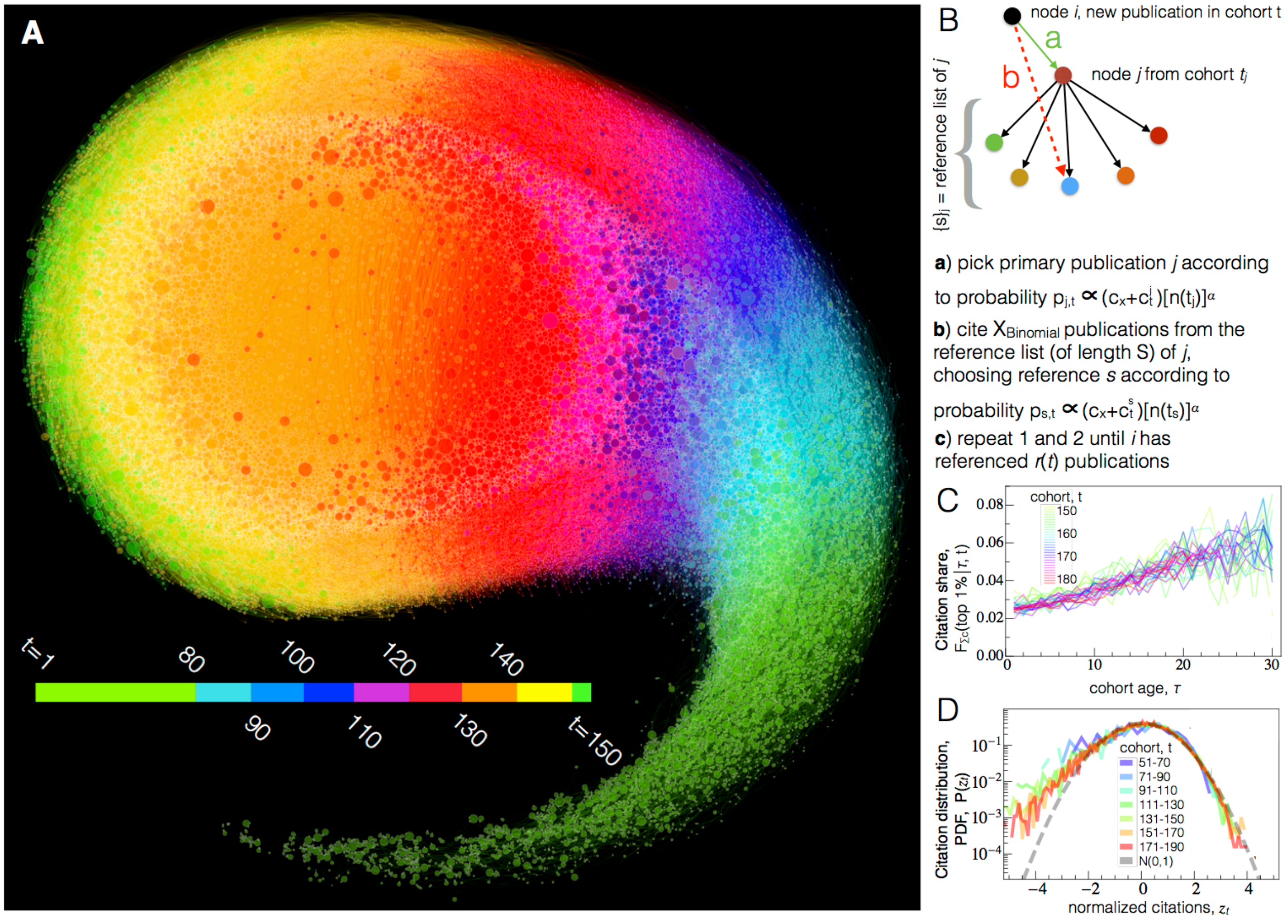}}
\caption{\label{modelschem}   {\bf The network of scientific citations and the  inflation at the knowledge frontier: model and empirical analysis.} (A) Visualization of the citation network produced by the citation model, emphasizing the long-term expansion of the knowledge stock juxtaposed by the relative thickening of the most recent layers. The latest publication cohort  represents a `growth cone'  which mediates the evolving connectivity of the growing network via the key redirection  process included in our model, wherein  more central and more recent publications are more likely to be referenced. The network was generated using  the parameters given in Fig. S6, and is comprised of  $N(T=150)=41,703$ nodes and $379,454$ links.  Here we show all the publications (nodes) from periods $t=[1,150]$ and assigned a color to each node  according to its age (see legend), e.g. nodes from  periods $t=[1,79]$  are colored green, nodes afterwards are colored according to decade, except for the nodes from $t=$150 (the only nodes from this decade) which are colored lime green. The relatively large size  of this last cohort, as compared to the first 149 periods, emphasizes the crowding out of the old by the new. In order to emphasize nodes from every cohort, the node size is proportional to the normalized citation count  $z^{p}_{t}$.
(B) The model comprises three complementary mechanisms: preferential attachment ($PA$), crowding out induced by growth, and the redirection of citations via reference lists. 
The ``crowding out'' ($n(t)$) and $PA$ ($c_{\times}+c^{j}_{t}$) factors in Eq. \ref{Pcj} balance each other, otherwise the citation rate decays rapidly after $1/(\alpha g_\text{R}) \approx 4$ periods or the first-mover advantage is overwhelmingly dominant.
However, obsolescence due to crowding out by new publications  can be overcome by the redirection process {\bf b)} operating through the reference list of an intermediate publication. The  model parameter $\beta$ controls the rate at which references occur via  {\bf b)} for every initial reference from  {\bf a)}, thereby capturing  the impact of shifts in ``hyperlinking'' citation behaviors. The model reproduces the various empirical trends, e.g.  (C)  the increasing  citation share $F_{\sum c}(1\%\vert t,\tau)$ of the top 1\% of publications from a given cohort $t$ \citep{Barabasi2012}, and  (D)  the log-normal distribution of citations received  \citep{UnivCite}. }
\end{figure*}

More specifically, each new publication $i$ from cohort $t$ makes $r(t)$ references by two distinct processes illustrated in the schematic  Fig. \ref{modelschem}(B): 
\begin{enumerate}[(a)]
\item direct ``browsing'' citation of a primary source  publication $j$,  and 
\item redirection to other articles appearing in the reference list of $j$. 
\end{enumerate}
 The relative rates of (a) and (b) are controlled by a parameter $\beta = \lambda/(\lambda+1)$, which is used in step (b)  to choose the random number $x$ of references cited from the reference list of $j$, where $x$ is drawn from a Binomial distribution with mean $\langle x \rangle = \lambda$.
 For any given simulation, let $r^{b}(t)$ be the total number of references occurring via process (b) in $t$, then on average $r^{b}(t) \approx \beta r(t)$, with the small discrepancy for small $t$ arising from finite size effects due to fixed upper limit in the reference list length, $r(t)$.
 
Figure \ref{modelschem}(A) shows a network visualization of a single realization of our model, and  emphasizes the empirical 12-year doubling period of the number of references produced -- the entire history of the growing system is rather quickly  overshadowed by  the new batch of publications/references from merely the most recent period!
  For this network of size $N(T=150)=41,703$ nodes and $L=379,454$ links, we calculate a modularity  $= 0.208$ and a mean clustering coefficient $= 0.018$ indicating the relatively high rate of triadic closure (i.e. since 0.018 $\gg L/(N(N-1)/2)=10^{-4}$, this means that the clustering coefficient is relatively large considering the number of edges and nodes).

 Our model  captures  several important features of the science  citation network.
First, it  incorporates the exponential  growth of the system, both in  $n(t)$ as well as  $r(t)$, a feature which is not taken into account in recent citation models which assume that the citation sources produce a constant number of references per time unit \citep{peterson_nonuniversal_2010,Golosovsky_stochastic_2012,wang_quantifying_2013,Golosovsky_uncovering_2014}.

Second, while we implement classic linear  preferential attachment ($PA$) \citep{Simon_class_1955,Barabasi_evolution_2002,Jeong_Measuring_2003, Redner2005PA,peterson_nonuniversal_2010}  in the link creation probabilities,
we also include an additional obsolescence term that captures the crowding-out effect induced by the growth of the system.
Combined, the attachment (citation) probability  of node (publication)  $j$ from $t_{j}$ is proportional to the weight
\begin{equation}
\mathcal{P}_{j,t} = (c_{\times}+c^{j}_{t})[n(t_{j})]^{\alpha} \ ,
\label{Pcj}
\end{equation}
where $c^{j}_{t}$ is the total number of citations received  by $j$ up to $t$, $n(t_{j})$ is the number of new nodes entering in period $t_{j}$, $\alpha \equiv 5$ is a scaling parameter controlling the characteristic obsolescence rate, and $c_{\times} \equiv 7$ is a citation  threshold, above which preferential attachment ``turns on''.  
A recent study has shown evidence for $c_{\times}$ on the order of 1  \citep{Golosovsky_transition_2013}, and in a general analysis of network models, this offset parameter is supported against alternative models \citep{Medo_statistical_2014}. 
Here  we find that $c_{\times} =7$ provides the best matching of the model and the empirical data with  respect to the $P(\Delta_{r})$ distributions shown in Fig. \ref{DistDr}.

The obsolescence factor $[n(t_{j})]^{\alpha}$ counteracts $PA$, because for two nodes with the same citation tally, the newer node will be preferentially selected -- i.e. ``crowding out'' of the old by the new. As a result, the relative attachment rate between any two given publications with the same number of citations but with a difference in age, $t_{j'} \geq t_{j}$, is given by $\mathcal{P}_{j,t} /\mathcal{P}_{j',t'}  = \exp[-\alpha g_{n} (t'-t)]$.
Thus, the crowding out provides a mechanistic explanation for the  decaying obsolescence (attention) factor. In addition to preferential attachment, this feature of our model is necessary  in order  to reproduce stylized facts associated with obsolescence in real citation networks \citep{medo_temporal_2011}.

And third, following the  citation of  $j$,  a random number of publications from  the reference list of $j$ are also cited.
This redirection step provides a ``backdoor''  to overcome the obsolescence induced by the growth of the system since articles in the reference list are likely to be older.

In addition to reproducing the trends  in  $F(c\leq C \vert t)$, $G(t)$, and $P(\Delta_{r})$, our model reproduces numerous other stylized facts representing  both static and dynamic features of the  empirical citation network. First,  the mean citation lifecycle is found to decay exponentially \citep{parolo_attention_2015}.  Second, there is a relatively high mean clustering coefficient typical of real  citation networks \citep{Holme_Growing_2002,ren_modeling_2012}. Third,  we demonstrate that our model reproduces the increasing citation share of the top cited percentile of publications \citep{Barabasi2012}, see Fig.  \ref{modelschem}(C).
And fourth,  we demonstrate that  normalizing citation counts within age cohort according to the logarithmic transform
\begin{equation}
z^{p}_{t}= (\log (c^{p}_{t})-\mu_{LN,t})/\sigma_{LN,t} \ ,
\end{equation} 
where  $\mu_{LN,t}=\langle \log (c^{p}_{t}) \rangle$ and $\sigma_{LN,t}= \sigma[ \log (c^{p}_{t})]$ are the  mean and the standard deviation of the logarithm of $c^{p}_{t}$ calculated across all $p$ within each  $t$,  results in a   citation distribution $P(z)$ that  is well-fit by the log-normal  distribution \citep{UnivCite}, see Fig.  \ref{modelschem}(D). In Fig. \ref{modelschem}(A) we implemented  a visualization scheme where each node size is proportional to the normalized citation count $z^{p}$,  and as a result, there is no visible temporal bias in the size distribution of the nodes.  
In this way, we demonstrate how an appropriate normalization that leverages the underlying statistical regularities of the data generating process can be useful for cross-temporal comparison.

We discuss these empirical benchmarks in further detail in the SM text and Figs. S6-S7, where we leverage the generative capacity of the citation network model to explore the effects of sudden perturbations of the system parameters during the network's evolution.
In the remainder of this section, we discuss the two most relevant perturbations.

\begin{figure*}
\centering{\includegraphics[width=0.99\textwidth]{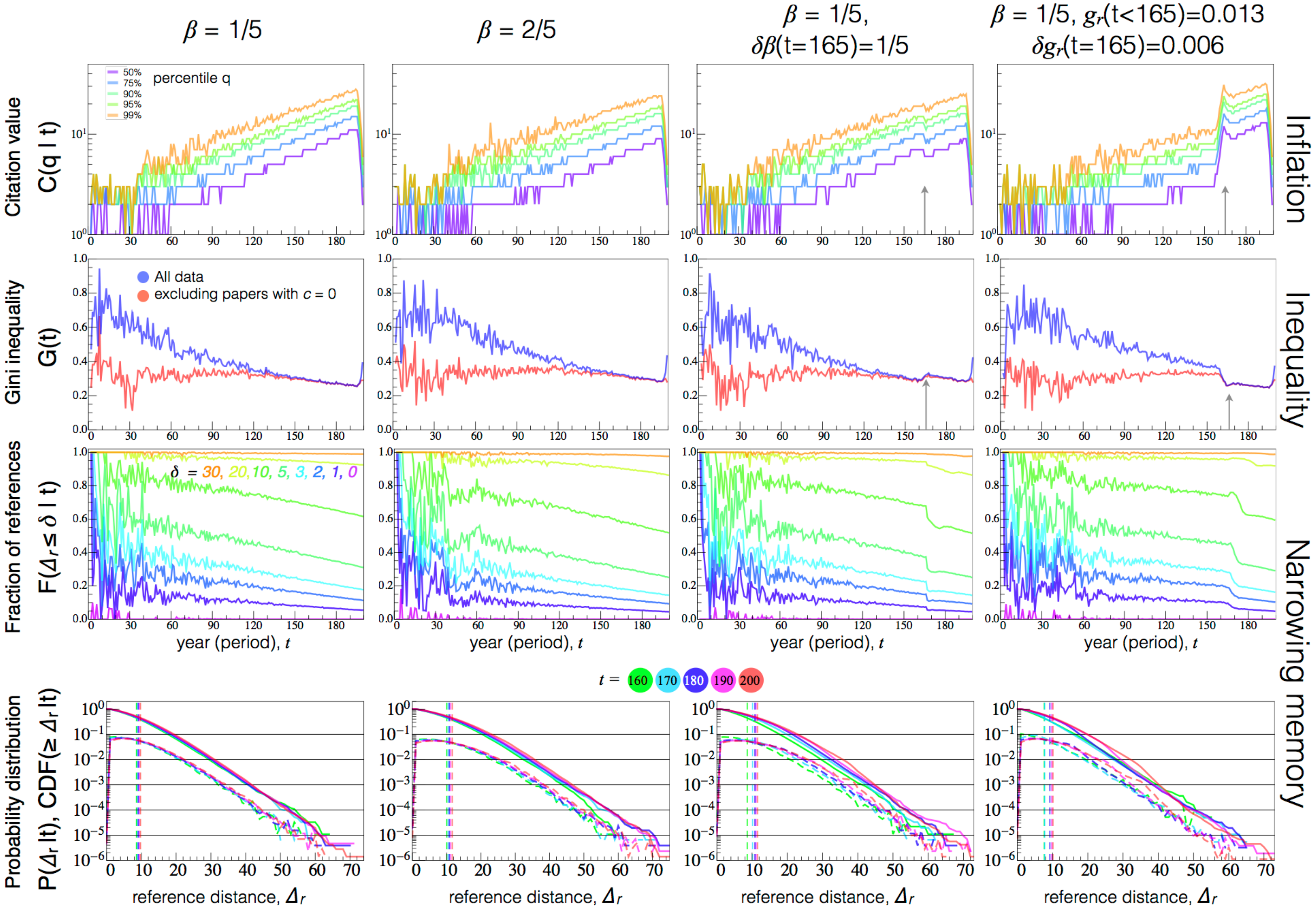}}
\caption{\label{modelcompare}  {\bf Monte-Carlo simulation of the science citation network: with the redirection mechanism ($\beta > 0$).} 
Each column represents a different modeling parameter set: the first and second columns differ only in the $\beta$ value; the third column represents a perturbation at $t=165$ from $\beta =1/5$ to $2/5$; and the final column represents the scenario with $\beta=1/5$ where the reference list growth rate $g_{r}$ is boosted from $0.013$ to $0.019$ at $t=165$. (First row) Inflation demonstrated by the persistent exponential growth of the  citation value $C(q \vert t)$ corresponding to the quantile $q$ indicated in the plot legend. For example, the citation value corresponding to the 99th percentile grows from roughly 3 in $t=30$ to 30 in $t=195$ for the unperturbed simulations   with $\beta=1/5$ and $\beta=2/5$.  (Second row) The Gini index $G(t)$ of the total number of citations after 5 years from publication measures the citation inequality of the citation distribution. The model also indicates that the decreasing inequality is largely due to the decreasing proportion of uncited publications. Nevertheless, for large $t$ the fraction of uncited publications is approximately zero, and so the decline in $G(t)$ is also induced by the growth of the system.  (Third row)  The fraction $F(\Delta_{r} \leq \delta \vert t)$ of references from year $t$ going to publications within the interval $[t-\delta,t]$ shows nonlinear behavior in the perturbed systems, with sharp declines indicating that the perturbation causes a significant fraction of references to be directed back  further than $\delta$ years in the past. 
These results capture the complex effects of  growing sources of references and the subsequent crowding out of old publications by new publications. 
(Fourth row) The cumulative distribution $P(\geq \Delta_{r}\vert t)$ (solid lines) and the probability density function $P(\Delta_{r}\vert t)$ (dashed lines) of  reference distance $\Delta_{r}$, conditional on the publication cohort $t$. Vertical lines indicate the mean of each conditional distribution for varying $t$. To improve the data size, each $P(\Delta_{r}\vert t)$ and $CDF(\geq \Delta_{r}\vert t)$ are calculated by pooling the reference data from the 3-period interval $[t-2,t]$. For example, the scenario with constant $\beta =1/5$ shows that medium $\Delta_{r}$ values becoming  more frequent for $t>160$. At the same time, recent publications corresponding to $\Delta_{r}\lesssim 4$ are being cited less and less. 
 Other parameters used for each simulation are $T\equiv$  200  MC periods ($\sim$ years), $n(0)$ = 10 initial publications, $r(0)\equiv$ 1 initial references, exponential growth rates $g_{n} \equiv 0.033$ and $g_{r} \equiv 0.018$ ($g_{R}=g_{n}+g_{r}$), citation offset $C_{\times}=6$, and life-cycle decay factor $\alpha \equiv 5$ so that $1/(\alpha g_\text{R}) \approx 4$ periods.}
\end{figure*}

First, we tested the synthetic citation network's response to a sudden increase in the parameter 
\begin{equation} 
\beta = r^{b}(t)/r(t) \ ,
\end{equation}
 which controls the rate of redirected referencing  (``hyperlinking'') via the redirection step (b) illustrated in Fig \ref{modelschem}. The sudden perturbation  $\beta  =0.2 \rightarrow 0.4$ at $t=165$ (see Fig. \ref{modelcompare} -- panel column 3) is meant to represent a shift from 1 in 5 citations to 2 in 5 citations occurring via mechanism (b), thereby simulating the  sudden  emergence of online journals.
This perturbation tests the conclusion of  Evans' study \citep{Evans18072008}, which used an econometric regression approach to determine the  impact of online journal archives on citation patterns, the results of which indicated that a negative relation between reference distance and online journal availability was due to a socio-technological shift captured by the emergence of   ``hyperlinks'', which facilitate the way in which researchers follow the reference trail from article to article.
However, in contrast to Evans' finding of a decreasing $\Delta_{r}$ due to the paradigm shift in hyperlinking, our model indicates that increasing the strength of the redirection process decreases {\it both} the frequency of $\Delta_{r} < \Delta^{-}_{r}$ and $\Delta_{r}>\Delta^{+}$, consistent with the empirical shifts in $P(\Delta_{r})$. Thus, our model suggests that the hyperlink effect  uncovered by Evan  is likely  relatively small compared to  the counter-effect induced by the growth of the system. Also, this perturbation causes a decrease in $C(q \vert t)$ and an increase in $G(t)$, because more references are redirected to older publications as demonstrated by the shifts in $F(c\leq C \vert t)$, $P(\Delta_{r} \vert t)$ and $CDF(\geq \Delta_{r} \vert t)$ towards $\Delta_{r}$ in the intermediate range $[\Delta^{-}_{r},\Delta^{+}_{r}] \approx [8,45]$.

Second, we  simulated a perturbation representing a sudden increase in the growth rate of the reference list growth rate,  $g_{r} \rightarrow g_{r} + \delta g_{r} $, by suddenly increasing $g_{r}$ from 0.013 to 0.019  at  $t^{*} =165$ (see Fig. \ref{modelcompare} -- panel column 4). For comparison, a similar perturbation is shown in the third panel column of Fig. S7 which is simulated using $\beta$=0. These perturbations are intended to explore the impact of online journals which are  less likely to have stringent article length and reference list length requirements. As far as the  $P(\Delta_{r})$ distribution, we observe that this second perturbation to $g_{r}$ has the same qualitative impact as the first perturbation to $\beta$. Interestingly, however, the perturbation to $\beta$ increased the citation inequality $G(t)$ whereas the perturbation to $g_{r}$ decreased the citation inequality $G(t)$.\\


\noindent{\bf Inflation-corrected productivity and impact measures.} 
The growth of science affects different units of analysis in different ways. For publications, the growth reduces the relative visibility of an individual article within its publication (age) cohort. For careers, it affects the net result of tallying up citation counts over long periods, 
because combining year-specific  {\it nominal values} (or raw citation counts) can produce a drastically  different tally than the corresponding tally of {\it real  values} (deflated values). 

To provide a concrete example of how growth affects the quantitative  evaluation of research careers, in this section we analyze the productivity and citation impact of 551 research profiles -- the set of all publications of researcher $i$ -- for 190 biologists and  361 physicists, each with an $h$-index \cite{hirsch_index_2005} of 10 or greater, and with first publication year denoted by $y_{0,i}$. To be specific, for each profile $i$, we  then collected the longitudinal citation count $\Delta c^{p}_{t}$ measuring the number of citations arriving in year $t$ to each publication $p$ within their publication profile, and analyzed this set of  $N_{i}$  publications through 2010 for each $i$. 

 Because a reference list can only include any given publication once,  the total number of citations a publication $p$ could receive from all new publications from  year $t$ is upper-bounded by $n(t)$. 
\footnotetext[7]{Interestingly, the impact of the growth of science on the obsolescence of literature  was  previously considered  by Nakamoto who showed that upon normalizing citations by $R(t)$, the forward and backwards reference difference distributions are quite similar \citep{Nakamoto_synchronous_1988}, however without this normalization there are more apparent differences \citep{Glanzel_forwardbackward_2004}.}
Since there is  cross and intra-disciplinary variation in the growth of $n(t)$, we generalize this publication rate to  the number of publications produced in research area $a$ given by $n_{a}(t)$, using the TR disciplinary classification system to define the number of articles   relevant to  each set of biology and physics researchers in  given year $t$ (see \citep{petersen_reputation_2014} for further details of the $n_{a}(t)$ deflator used for biology and physics).  As such, the citation ``opportunity share''  $\Delta c^{p}_{t}/n_{a}(t)$  is an intuitive way of normalizing the number of citations received by $p$ in $t$.  

Hence, for each publication $p$ we denote $c^{p}_{T} = \sum_{t=0}^{T}\Delta c^{p}_{t}$ as the total number of citations  up to year $T \equiv 2010$ to  a given $p$ belonging to  $i$.
Then using these final citation tallies, for each  $i$ we calculated  the   $h$-index $h_{i}$ (a productivity measure) and the net citations $C_{i} = \sum_{p \in i} c^{p}_{T}$ (a net citation impact measure) \citep{hirsch_index_2005,petersen_statistical_2011,petersen_z-index:_2013}.  
We then recalculated  the deflated h-index $h^{D}_{i}$ and the deflated net citations 
\begin{equation}
C^{D}_{i} = \sum_{p\in i} s^{p}_{t}
\end{equation} 
for  each research profile $i$ using the deflated citation values,  
\begin{equation}
s^{p}_{T} = \sum_{t=0}^{T}\Delta s^{p}_{t}  \ \ \text{with} \ \ \Delta s^{p}_{t}  \equiv \Delta c^{p}_{t}\frac{n_{a}(2010)}{n_{a}(t)} , 
\label{SDeflatedC}
\end{equation} in each step of the citation tally process, choosing 2010 as the arbitrary baseline comparison year.
\footnotetext[8]{As in the measurement of the gross domestic product (GDP) of countries, it is important to distinguish the nominal rate of  growth (not accounting for inflation) and the real rate of growth (accounting for inflation).
As such, the citations are measured in terms of their relative value in the (arbitrary) base year $t\equiv 2010$, thereby accounting for the  growth of $n_{a}(t)$. Because of the generality of the approach, it can be extended to other units of analysis, e.g. institutions and countries, and various other socio-economic contexts. By way of analogy, in order to   compare player achievements/records across era,  deflation indices have recently been implemented  in the context of professional baseball in order to develop career achievement metrics that account for  inflation in the total number of possible player opportunities per season \citep{petersen_methods_2011}.}


We  separated the researcher profiles into age cohort subsets, depending on the career birth year, $y_{0,i}$, and measured the
difference between the traditional $h$-index $h_{i}$ and total citations $C_{i}$ and their deflated counterparts, $h_{i}^{D}$ and $C_{i}^{D}$.
Because each discipline includes 100 highly cited scientists, the range of $h_{i}$ and $C_{i}$ is rather broad, representing early career researchers with $h_{i}\sim10$ up to eminent scientists with $h_{i}>100$. The highly cited scientists are mostly from relatively early age cohorts, having started their careers between 1940 and 1969. 

\begin{figure*}
\centering{\includegraphics[width=0.99\textwidth]{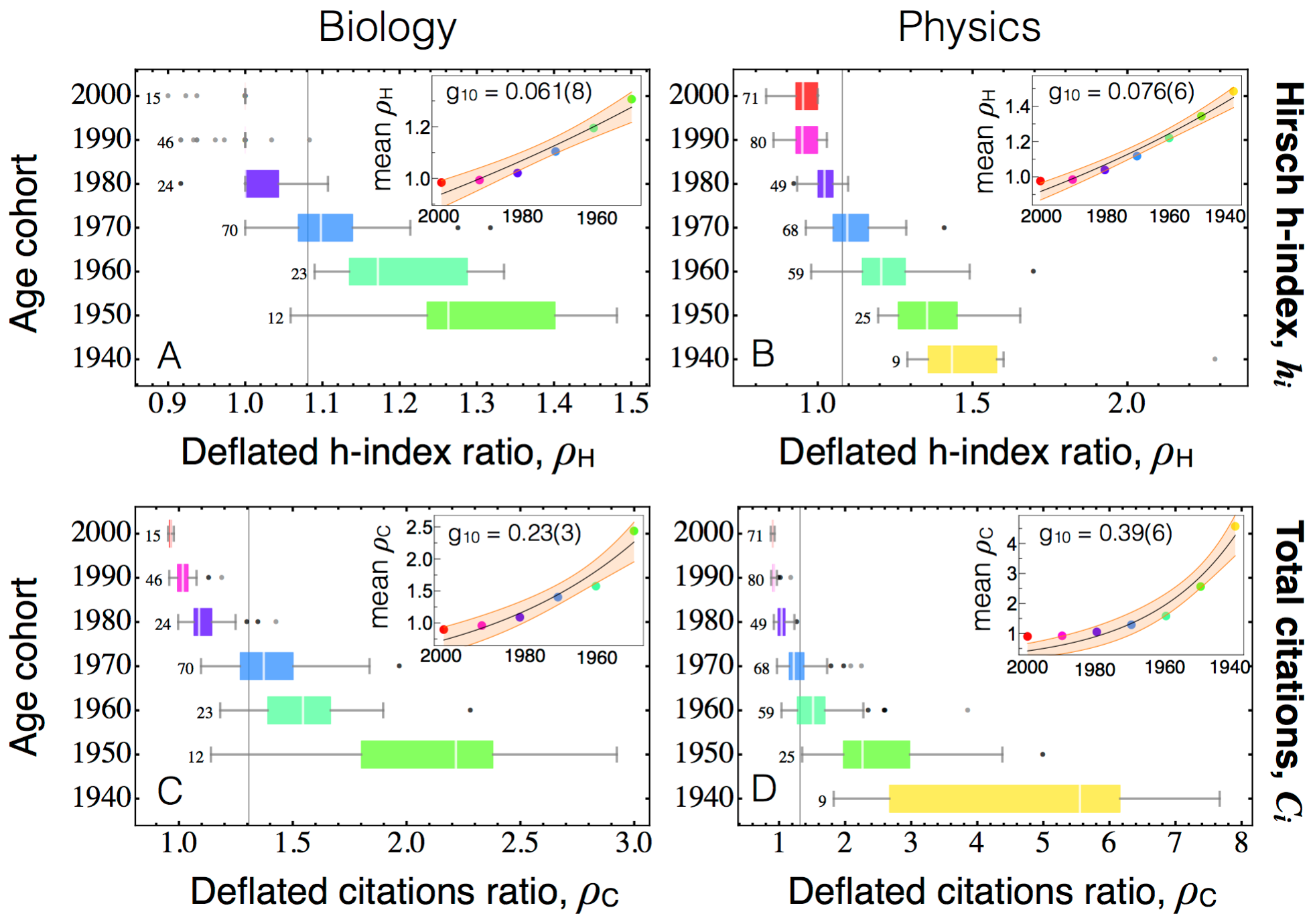}}
\caption{\label{hCdetrended}  {\bf Deflated productivity and impact measures by discipline and age cohort.}
The deflated $h-index$ ratio $\rho_{H,i}\equiv h^{D}_{i}/h_{i}$ and the deflated total citations ratio $\rho_{C,i}\equiv C^{D}_{i}/C_{i}$ indicate the factor increase that a researcher would receive if the deflated citation measure $s_{p}(T)$  (see Eq. \ref{SDeflatedC})  is used instead of $c^{p}(T)$ in calculating $h_{i}$ and $C_{i}$. Since we have used 2010 as the baseline year, researchers from the most recent cohorts have ratio values close to unity, whereas researchers from earlier cohorts by in large receive a relative boost $(r>1)$.
Shown are box-whisker  distributions of the deflated  $h$-index ratio $\rho_{H}$ (A,B)   and deflated citations ratio $\rho_{C}$ (C,D) by age cohort, with the midpoint of each box representing the median value; mean value across all data indicated by vertical  black line. 
The mean deflated  $h$-index ratio  value is $\langle \rho_{H}\rangle= 1.08$ (for both biology and physics). 
The mean deflated citations ratio  value is  $\langle \rho_{C}\rangle= 1.31$ (biology) and  $1.32$ (physics).
(insets) Progression of the mean $\rho_{H}$ and $\rho_{C}$ by each 10-year cohort, and the 90\% confidence interval indicated by the shaded region. 
Also listed are the estimates of the 10-year exponential growth factor, $g_{10}$, as defined in Eq. \ref{EqRfit}; the standard error in the last digit is indicated in parenthesis.}
\end{figure*}

We then analyzed  the ratios $\rho_{H,i} \equiv h^{D}_{i}/h_{i}$ and $\rho_{C,i} \equiv C^{D}_{i}/C_{i}$, which measure the relative impact of our deflating scheme on each individual's research profile. The mean  values calculated independent of age cohort,  $\langle \rho_{H} \rangle =1.08$ and $\langle \rho_{C} \rangle =1.31$ independent of discipline, are both greater than unity. However, by separating the $\rho_{H,i}$ and  $\rho_{C,i}$ values by age cohort,  Fig. \ref{hCdetrended} shows that remarkably high levels of ``underestimation'' can occur when  citation counts of older researchers are not adjusted for inflation.  

We used the trend in $\rho_{H}(t)$ and $\rho_{C}(t)$  to estimate  the overall inflation rate of the science achievement measures themselves by first calculating the mean value $\langle \rho(t) \rangle$ for  the researchers from each 10-year group (e.g. from the 2000s to the 1940s), which are plotted in the insets of Fig. \ref{hCdetrended}. We then used  the functional form
\begin{equation}
\langle \rho(t) \rangle=\rho_{0}\exp[g_{10}(2000-t)/10]
\label{EqRfit} 
\end{equation}
 to estimate the 10-year growth rate $g_{10}$. 
For $\rho_{H}$, we estimate $g_{10}\approx 0.061$ (biology) and $g_{10}\approx 0.076$ (physics), which means that for every 10 years in the past, $\rho_{H}$ grows by roughy 6 to 7\%.
For $\rho_{C}$, the percent growth is significantly larger, estimating $g_{10}\approx 0.23$ (biology) and $g_{10}\approx 0.39$ (physics).
Together these numbers quantify the  extent to which a 10-year time difference can alter the relative values of productivity and impact measures when one accounts for citation inflation. As a robustness check, we also pooled the individual $\rho_{H,i}$ ($\rho_{C,i}$) values along with each researcher's individual $y_{0,i}$ and estimated $g_{10}$, which resulted in nearly identical values.

\vspace{-0.2in}
\section*{Discussion}
\vspace{-0.2in}
\noindent{\bf What are the sources of scientific output inflation?} 
It is important to consider the main factors underlying the growth of science.
First, the  4\% annual growth of $n(t)$ is largely due to the growth of the  researcher population size, which over the period 1997-2012,  has been growing  at an average annual pace of 4-6\%    (see Fig. \ref{RnDInputGrowth}).  
It is important to note that these average rates neglect the heterogeneity within and across disciplines. Indeed,  the researcher population size, the amount of funding, and the subsequent number of scientific publications are highly dependent on the nuances of technological and institutional factors \citep{stokes_pasteurs_1997,stephan_how_2012}.

By way of example, consider  ``stem cell'' research, which  currently sits at the knowledge frontier, poised with the capacity for socio-economic and technological impact. This research field is characterized by 17\% annual growth following a considerable growth burst in the late 1980s due to singular discoveries and directed funding initiatives (see Fig. S1). Also important in driving the growth of $n(t)$ are paradigm shifts  such as academic word processors (LaTex), bibliographic organization tools, online submission and editorial services, and the advent of rapid open-access online publishing,    which  have facilitated the increasing pace of the publication process \citep{lariviere_long-term_2008}. 

As another factor driving inflation, consider  the sole contribution by the ``rapid'' open-access journal PLoS One, which grew over its first 6 years at an annual rate of 58\%, corresponding to a doubling rate of 1.2 years. To place this  growth in real terms, in 2012, after just 5 years since its inception, PLoS One represented roughly 1 out of every 1000 Science publications per year.  As such, the proliferation of  ``e-journals'' may have contributed to the perception that everything is publishable, thereby leading to a ``democratization of scientific credit''  -- the potential analog of a  market bubble. 

The referencing of prior literature produces an intergenerational transfer of `citation credit'. As such,  inflation in the production of references  and an inhomogeneous distribution of reference distance,  may together negatively impact the efficiency of the knowledge network, i.e. causing a retroactive ``crowding-out''  of prior literature. 
As illustrated in the network visualization in Fig. \ref{modelschem}, the deluge of recent literature may collectively reduce the  ability of scientists to explore the more distant corners of the knowledge network, thereby  decreasing the efficiency of scientific progress. 
Thus, it is important to quantify and understand trends in  science memory in order  avoid, among other systemic inefficiencies, the syndrome of `reinventing the wheel'.

This  narrowing of scientific scope, as depicted by the ``expertise field of view'' ($\theta$) in Fig. \ref{scienceschem}, may follow from various factors.
First, the training of scientists has become more focused on   specialization during the doctoral training, often to prepare for careers in large laboratory environments, together marking the end of the  solo ``renaissance'' genius era \citep{simonton_after_2013}. Second, the deluge of new literature may push researchers  to the limits of their individual cognitive abilities to browse and follow-up on all the  new and foundational  literature. As such,  within the exploratory space of reference lists,  the channels to distant literature may become increasingly narrow relative to the channels to more contemporary literature. Lastly, there may be narrowing due to the focusing power of circles of social influence arising, among others factors, from the increasing prevalence of  scientific reputation and journals associated with academic societies,  which may affect the publication and referencing process.

Nevertheless, handling the overwhelming volume of knowledge required to make scientific advancement may, in part, be overcome by the division of labor. Indeed, the number of coauthors per publication, a proxy for team size,  has also shown a persistent 4\% annual growth over the last half century in the natural sciences \citep{petersen_quantitative_2014}. 
Moreover, exploring the knowledge network is also facilitated by new  technologies for accessing, crowdsourcing (e.g. {\it Wikipedia.org}), searching, retrieving, storing,  and organizing knowledge \citep{Sparrow05082011}.\\

\noindent{\bf Contributions to the literature.}  We addressed two  outstanding disagreements in the literature concerning (a) the level of citation inequality across publications
 \citep{Evans18072008,acharya_rise_2014,petersen_inequality_2014,lariviere_decline_2009} and (b) the obsolescence of knowledge \citep{Evans18072008,wallace_small_2012,verstak_shoulders_2014}. We provided  clarity on these two issues by performing an in-depth analysis of  (i) the entire citation distribution, as measured  by the  Gini coefficient $G(t)$, and (ii)
the entire range of reference distances, quantified by the probability distribution $P(\Delta_{r})$.  In the case of $G(t)$, we found that the decreasing inequality in the number of citations received is largely due to the simultaneous decreasing trend in uncitedness.
In the  case of $P(\Delta_{r})$, we found that the trends  are rather nuanced,  and thus susceptible to misinterpretation if simple biased summary statistics are used. We found that the  fraction of references to literature of intermediate $\Delta_{r}$ range is increasing, accompanied by a decline in the attention to both very recent and very distant literature. 
More specifically, the largest decline in $P(\Delta_{r} \vert t)$ was  in the range $\Delta_{r}\leq 5$  years. 

These trends may reflect social factors associated with the rapid growth of $n(t)$. Instead of reading every new publication, researchers may increasingly depend on the ``wisdom of the crowd'' to collectively crowdsource the quality  of research -- i.e. as proxied by citation counts --  an evaluation  process which can take  several years to accumulate, thereby slowing down the ``digestion rate'' of academic literature. This is in part reflected by the citation threshold $c_{\times}$ which controls the onset of preferential attachment. Indeed, preferential attachment is a collective effect since it requires global information.

\begin{figure*}
\centering{\includegraphics[width=0.99\textwidth]{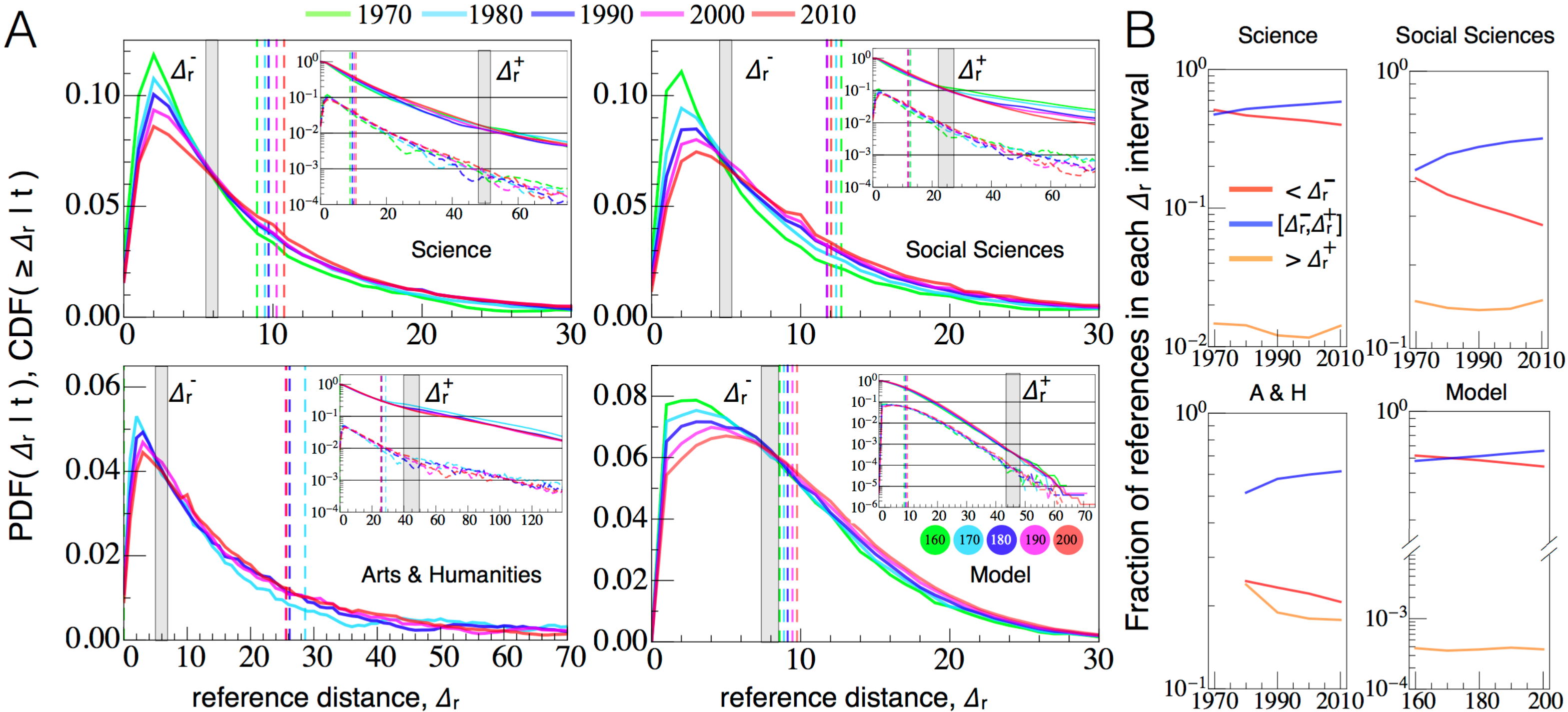}}
\caption{\label{DistDr}  {\bf Reduction of scientific attention in the near and far fields.} (A) Probability distributions $P(\Delta_{r} \vert t)$ for select $t$ indicated in the color legend. (inset)  $CDF(\geq \Delta_{r} \vert t)$ (solid curve) and probability distributions $P(\Delta_{r} \vert t)$ (dashed curve) on log-linear scale to emphasize the shifts in the distributions for large $\Delta_{r}$.  Each panel shows a small $\Delta_{r}^{-}$ regime for which  the  $P(\Delta_{r} \vert t)$  cross -- independent of $t$ -- signaling a  ``universal memory scale'' (fixed point) in the reference distance distribution: 
$\Delta^{-}_{r}\approx 6$ years (Sci.), 5 y (Soc. Sci.), 6 y (A\&H), and 8 y (Model). A second crossing point $\Delta^{+}_{r}$  is indicated in the empirical $CDF(\geq \Delta_{r} \vert t)$, such that  the fraction of citations going to  publications with $\Delta_{r}>\Delta^{+}_{r}$ is decreasing for larger $t$: $\Delta^{+}_{r}\approx 50$ years (Sci.),  20 y (Soc. Sci.), 40 y (A\&H), and 45 y (Model). Interestingly, Science exhibits an increasing mean value with time, whereas Soc. Sci. and A\&H indicate a decreasing mean value, demonstrating how single-value distribution measures can yield misleading comparisons. The lower-right panel shows the results of our model;   
 Model parameters are listed in  Fig. S6. (B) The narrowing range of scientific attention, demonstrated by the  fraction of references in 3 non-overlapping intervals, $\Delta_{r}<\Delta_{r}^{-}$, $\Delta_{r}^{-} \leq \Delta_{r} \leq \Delta_{r}^{+}$, and $\Delta_{r}>\Delta_{r}^{+}$. 
}
\end{figure*}

We also found that the fraction of reference distances greater than $\Delta_{r}^{+}$, measured by  $CDF(\geq \Delta_{r}^{+}\vert t)$, is also decreasing over time for each  subject area, with a  recent upturn in the last 10 years  consistent with an increasing redirection effect (see Fig. \ref{DistDr}B). The crossover value varies by subject area with $\Delta^{+}_{r}\approx 50$ y (Science),  $\Delta^{+}_{r}\approx 20$ y (Social Sciences) and $\Delta^{+}_{r}\approx 40$ y (A\&H). This decline in attention to distant literature (decreasing hyperopia) appears at first to be in disagreement with  results following the same method  \citep{wallace_small_2012,verstak_shoulders_2014}. However the discrepancy is likely due to the fact that these analyses only investigated the citation trends for  $\Delta_{r}\leq 10$, 15, and 20 years, and so they did not investigate sufficiently large $\Delta_{r}$ to access the trends in the truly classic literature. 

Also, because the  $P(\Delta_{r})$ distribution is right-skewed with a heavy tail, analyzing  only select  summary statistics -- such as the mean or cumulative fractions -- can be misleading, as demonstrated by the fact that the  mean reference distance $\langle \Delta_{r} \vert t \rangle$ is increasing for Science but decreasing for Social Sciences and A\&H (see Fig. \ref{DistDr}A). For this reason, we analyzed the entire distribution captured by $P(\Delta_{r})$ and $CDF(\geq \Delta_{r})$.
The decline in $CDF(\geq \Delta_{r}^{+}\vert t)$ is likely due to technological turnover and the emergence of new disciplines which have relatively few foundational publications to reference - a  branching process of innovation that is not captured by the model. The content of  academic literature also evolves differently in A\&H versus Science: in the former, references are often historical or artifactual in context, referring to quasi-static representations, as opposed to the  dynamic concepts and methods that can rapidly evolve in the latter. 

A particularly interesting feature of the shifts in $P(\Delta_{r} \vert t)$ is the constant fixed  point $P(\Delta_{r}^{-} \vert t)$, which appears to be independent of $t$. Thus, because of this stability,  the reference distance $\Delta_{r}^{-} \approx 6$ y  can be used to classify knowledge as recent ($\Delta_{r}\leq \Delta_{r}^{-}$) or contemporary ($\Delta_{r}^{-} \leq \Delta_{r} \leq \Delta_{r}^{+}$), representing a fundamental  time scale characterizing the advancement of the scientific endeavor. Similarly, although less precise, a second  crossing point  $\Delta_{r}^{+}$ in the cumulative distribution $CDF(\geq \Delta_{r})$ distinguishes  the classic literature   ($\Delta_{r} \geq  \Delta_{r}^{+}$). Fig. \ref{DistDr}B shows that the fraction of references distances falling into the range $[\Delta_{r}^{-},\Delta_{r}^{+}]$ has steadily increased, growing by roughly 24\% in Science, 30\% Soc. Sci., and 19\% in A\&H over the last 50 years.\\

\noindent{\bf A new benchmarked growth model for the evolution of the knowledge network.} To further investigate the impact of these paradigm shifts, we developed a generative citation model which we validated by  reproducing numerous empirical features of the science citation network. Despite the model's success, it has some clear limitations which are worth mentioning. First, we do not incorporate the intrinsic  quality of new publications nor any other node   features  (e.g. journal, authors, author affiliations) meaning that our model lacks heterogeneity in the  intrinsic fitness of each $p$, a factor which can explain the extremely  wide variation in long-term citation impact of individual publications \citep{wang_quantifying_2013}. 

Second, the model lacks social factors, such as author-specific effects such as reputation \citep{petersen_reputation_2014}, collaboration \citep{Barabasi_evolution_2002,petersen_quantitative_2014,petersen_quantifying_2015},  and self-citation \citep{wallace_small_2012}, which are inextricable  features of the science system that lead to important correlations in the coevolutionary dynamics  of the citation network. In particular, the tendency for authors to self-cite, in other words their heightened awareness of their own work, may affect the distribution of $\Delta_{r}$ in both the short and the long term. Indeed, by incorporating a social layer into our model, one which captures self-citation, we might be able to better match the model predictions   to the empirical  $P(\Delta_{r} \vert t)$ distributions in the small $\Delta_{r}$ regime.

Nevertheless, our primary goal was to use the model to determine the response of the synthetic science system to  four scenarios operationalized as modifications and  sudden perturbations of the model parameters:  
\begin{enumerate}[(i)]
\item no redirection mechanism ($\beta=0$), see Fig. S7;
\item a sudden perturbation increasing $\beta$ after  $t^{*}$, thereby increasing the frequency of the redirection process (b) relative to the direct process  (a), 
 representing a paradigm shift in ``hyperlinking''   \citep{Evans18072008}, see Fig. \ref{modelcompare};
 \item a sudden perturbation to $g_{r}$ representing a sudden increasing of reference list lengths after  $t^{*}$, see Fig. \ref{modelcompare};
 \item a sudden perturbation to $g_{n}$, causing either a decrease or increase in the  growth rate of the system after  $t^{*}$ (see Fig. S7).
\end{enumerate}
 We discuss these {\it in silico} experiments  in further detail in the SM text and   Fig. S7. 

In all, these MC simulations  provide  various insights into the evolution of the citation network that are not possible otherwise, since closed-form analytic methods are typically not tractable for such time-dependent network growth models.
First,  our simulations  indicate how $n(t)$,  the ``crowding-out'' factor  included  in Eq. \ref{Pcj}, can indirectly cause the    saturation (obsolescence) of the citation lifecycle. For a visual example, this ``crowding out'' is indicated in Fig. \ref{modelschem}(A) where the publications from the final period $t=150$ (bright yellow) are already prominent compared to the entire corpus  of the preceding 149 periods. Second, our model clarifies how the narrowing of attention to more recent literature   counteracts the positive feedback arising  from the preferential attachment, together producing  realistic citation (knowledge) life-cycles that peak and then decay exponentially after a time scale $\sim 1/(\alpha g_\text{R}) \approx 4$ periods (see Fig. S6). Third, redirected references represent a mechanism to  overcome the  implicit citation lifecycle  induced by the growth of $n(t)$ and $r(t)$, because the references that arise from the redirection process cite older literature on average. 
And finally, we  also find that  the growth of $n(t)$ is necessary for sustaining the citation of recent literature, as indicated by the perturbation $g_{n,t} \rightarrow 0$, which results in a sudden decline   in the number of   citations received and a sudden increase in the citation inequality  $G(t)$, as demonstrated in Fig. S7.\\

\vspace{-0.2in}
\section*{Conclusions}
\vspace{-0.2in}
\noindent We conclude with two policy recommendations. First, our results suggest that the implementation of citation deflator based upon the publication rate $n(t)$ should be used in evaluation scenarios involving the aggregation of citation counts across time and the comparison of citation counts between age cohorts. 
 Let us make our point by first referring to a  method that has been implemented in the literature to address the right-censoring bias when comparing the citation count of publications from 2 different years -- which is to count only the  citations accrued in the first $\Delta t$ years (e.g. here in subsections 4.1 and 4.2 we used the fixed window $\Delta t = 5$ y). In this way, both publications are effectively compared at the same age; nevertheless, we emphasize that this approach  {\it is not sufficient} to eliminate the inflation bias arising from the fact that there are more citations produced in the later $\Delta t$-year period than the prior $\Delta t$-year period. 
  Instead, a deflator index is needed to correct for this particular type of statistical bias. 

This recommendation is further anchored on  the fact that, in common practice, citations are summed with no variable weight given to the source year, which neglects the fact that more and more  references are produced each year, representing  a challenging measurement problem. However, because a  publication can only be cited once by any other publication, normalizing citations received in any given year $t$  by the publication rate $n_{a}(t)$ (within a restricted   research domain $a$) is an intuitive way to control for the underlying exponential growth of science.
Two recent studies, one focusing on  reputation growth of individual careers \citep{petersen_reputation_2014} and the other focusing  on the citation lifecycle of individual publications  \citep{parolo_attention_2015}, have  demonstrated the use of   $n_{a}(t)$ as a deflator index, effectively measuring time in units of contemporaneous output. 

Here we further  demonstrate the use of a deflator index 
in the calculation of  researchers' net productivity and citation impact, as captured by the $h$-index and total citations  $C_{i}$, respectively. 
Our approach involves a simple rescaling of citations according to an arbitrary baseline year $t_{b}$ using the deflated citation values given by
 $\Delta s^{p}_{t}  \equiv \Delta c^{p}_{t} \times [n_{a}(t_{b})/n_{a}(t)]$ using a standard baseline year $t_{b} \equiv 2010$ (see Eq. \ref{SDeflatedC}). 
Similar in-house methods for normalizing citation measures may be part of the existing ``best-practice'' within select research evaluation groups. However, we call for a more concerted effort in  the  `science of science' research community to use citation deflators.

As a particularly relevant case study, we show that  the current (standard)  method of aggregating citation counts can lead to an underestimation of  $h_{i}$ and $C_{i}$ with larger penalties on researchers from older age cohorts. Another way to appreciate this is to consider the seminal $h$-index paper, where Hirsch writes:   ``for [physics] faculty at major research universities,  $h \approx 12$  might be a typical value for advancement to tenure (associate professor) and that $h \approx 18$ might be a typical value for advancement to full professor'' \citep{hirsch_index_2005}.  While these numbers may have been reasonable roughy a decade ago,  our analysis shows why they are undervalued  by present day measure.  Using the rate of change in $\rho_{H}$ by each age cohort, our estimates indicate that these $h$-index thresholds should be increased by  6 to 7\% every decade -- i.e. each $h$-index value above should be increased by roughly 1. Similarly, if $C_{i}=10,000$ citations were a benchmark one decade, our estimates indicate that the benchmark should be increased  by 20\%-40\% the following decade, depending on the discipline.

However, obtaining the deflator time series $n_{a}(t)$ is not a straightforward process for any arbitrary discipline  ``$a$'' or combination thereof, and so we call on large citation indices (Google Scholar, TR, Scopus) to develop and provide researchers with a basic yet efficient search query interface for obtaining   $n_{a}(t)$ so that normalized citation measures can be more readily calculated.

Second, because the supply of references is a source of inflation, journals should consider standardizing the maximum number of references per publication, depending on the article page length or  type (letter, article, review, etc.). This would limit the growth in the total supply of references, $R(t)$, and may discourage other bad habits such as self-citation with the intent to surgically enhance one's $h$-index \cite{petersen_z-index:_2013}.

And finally, our model of citation dynamics provides the opportunity to test the impact of new growth regimes due to open-access publication on the science citation network. 
Models which provide insight into the interplay between citation measures, scientists citing behavior, and growth of scientific output, are an essential ingredient of our progress in quantitative evaluation of scientific production  \citep{Wildson_need_2015,TheMetricTide} and the science of science policy \citep{fealing_science_2011}. \\

\noindent {\bf Acknowledgments} 
The authors are grateful for helpful discussions with A.-L. Barab\'asi, A. Bonaccorsi and O. Penner. AMP and FP acknowledge financial support 
 from the Italian Ministry of Education, PNR Crisis Lab, \href{www.crisislab.it}{www.crisislab.it}. 
  The authors also acknowledge the opportunity to  receive feedback via  COST Action TD1210 ``KnowEscape.''
  Science funding  data is openly available from the World Bank \citep{WorldBankData}.   Certain data included herein are derived from the Science Citation Index Expanded, Social Science Citation Index and Arts \& Humanities Citation Index, prepared by Thomson Reuters, Philadelphia, Pennsylvania, USA, Copyright Thomson Reuters, 2011.




\begin{thebibliography}{10}

\bibitem{stokes_pasteurs_1997}
Stokes DE
\newblock (1997) \emph{Pasteur's Quadrant: Basic Science and Technological
  Innovation}
\newblock (Brookings Institution Press, Washington, {D.C.}, {USA}).

\bibitem{stephan_how_2012}
Stephan P
\newblock (2012) \emph{How Economics Shapes Science}
\newblock (Harvard University Press, Cambridge {MA}, {USA}).

\bibitem{lariviere_long-term_2008}
Lariviere V, Archambault E, Gingras Y
\newblock (2008) Long-term variations in the aging of scientific literature:
  {From} exponential growth to steady-state science (1900-2004).
\newblock \emph{JASIST} 59:288--296.

\bibitem{ASI:ASI20936}
Althouse BM, West JD, Bergstrom CT, Bergstrom T
\newblock (2009) Differences in impact factor across fields and over time.
\newblock \emph{JASIST} 60:27--34.

\bibitem{parolo_attention_2015}
Parolo PDB, {et~al.}
\newblock (2015) Attention decay in science.
\newblock \emph{Journal of Informetrics} 9:734 -- 745.

\bibitem{Franck_Scientific_1999}
Franck G
\newblock (1999) Scientific communication--a vanity fair?
\newblock \emph{Science} 286:53--55.

\bibitem{Klamer_Attention_2002}
Klamer A, van Dalen HP
\newblock (2002) Attention and the art of scientific publishing.
\newblock \emph{Journal of Economic Methodology} 9:289--315.

\bibitem{fealing_science_2011}
Fealing KH, {eds.}
\newblock (2011) \emph{The science of science policy: A handbook.}
\newblock (Stanford Business Books, Stanford {CA}, {USA}).

\bibitem{garfield_citation_1955}
Garfield E
\newblock (1955) Citation indexes for science: A new dimension in documentation
  through association of ideas.
\newblock \emph{Science} 122:108--111.

\bibitem{deSollaPrice_networks_1965}
de~Solla~Price DJ
\newblock (1965) Networks of scientific papers.
\newblock \emph{Science} 149:510--515.

\bibitem{100yearsphysics}
Sinatra R, Deville P, Szell M, Wang D, Barabasi AL
\newblock (2015) A century of physics.
\newblock \emph{Nat Phys} 11:791--796.

\bibitem{Sparrow05082011}
Sparrow B, Liu J, Wegner DM
\newblock (2011) Google effects on memory: Cognitive consequences of having
  information at our fingertips.
\newblock \emph{Science} 333:776--778.

\bibitem{schwartz_rise_1997}
Schwartz CA
\newblock (1997) The {Rise} and {Fall} of {Uncitedness}.
\newblock \emph{College \& Research Libraries} 58:19--29.

\bibitem{Wallace2009296}
Wallace ML, Lariviere V, Gingras Y
\newblock (2009) Modeling a century of citation distributions.
\newblock \emph{Journal of Informetrics} 3:296 -- 303.

\bibitem{lariviere_decline_2009}
Lariviere V, Gingras Y, Archambault E
\newblock (2009) The decline in the concentration of citations, 1900-2007.
\newblock \emph{JASIST} 60:858--862.

\bibitem{acharya_rise_2014}
Acharya A, {et~al.}
\newblock (2014) Rise of the {Rest}: {The} {Growing} {Impact} of {Non}-{Elite}
  {Journals}.
\newblock \emph{arXiv:1410.2217}.

\bibitem{petersen_inequality_2014}
Petersen AM, Penner O
\newblock (2014) Inequality and cumulative advantage in science careers: a case
  study of high-impact journals.
\newblock \emph{{EPJ} Data Science} 3:24.

\bibitem{Barabasi2012}
Barabasi AL, Song C, Wang D
\newblock (2012) Publishing: Handful of papers dominates citation.
\newblock \emph{Nature} 491:40.

\bibitem{Evans18072008}
Evans JA
\newblock (2008) Electronic publication and the narrowing of science and
  scholarship.
\newblock \emph{Science} 321:395--399.

\bibitem{wallace_small_2012}
Wallace ML, Lariviere V, Gingras Y
\newblock (2012) A {Small} {World} of {Citations}? {The} {Influence} of
  {Collaboration} {Networks} on {Citation} {Practices}.
\newblock \emph{PLoS ONE} 7:e33339.

\bibitem{verstak_shoulders_2014}
Verstak A, {et~al.}
\newblock (2014) On the {Shoulders} of {Giants}: {The} {Growing} {Impact} of
  {Older} {Articles}.
\newblock \emph{arXiv:1411.0275}.

\bibitem{Broad_publishinggame_1981}
Broad W
\newblock (1981) The publishing game: getting more for less.
\newblock \emph{Science} 211:1137--1139.

\bibitem{petersen_reputation_2014}
Petersen AM, {et~al.}
\newblock (2014) Reputation and impact in academic careers.
\newblock \emph{Proc. Natl. Acad. Sci. USA} 111:15316--15321.

\bibitem{Nakamoto_synchronous_1988}
Nakamoto H
\newblock (1988) in \emph{Informetrics 87/88: Select Proceedings of the 1st
  International Conference on Bibliometrics and Theoretical Aspects of
  Information Retrieval}, eds{} Egghe L, Rousseau R
\newblock (Elsevier, New York), pp 157--163.

\bibitem{Glanzel_forwardbackward_2004}
Gl{\"a}nzel W
\newblock (2004) Towards a model for diachronous and synchronous citation
  analyses.
\newblock \emph{Scientometrics} 60:511--522.

\bibitem{Holme_Growing_2002}
Holme P, Kim BJ
\newblock (2002) Growing scale-free networks with tunable clustering.
\newblock \emph{Phys. Rev. E} 65:026107.

\bibitem{vazquez_growing_2003}
Vazquez A
\newblock (2003) Growing network with local rules: Preferential attachment,
  clustering hierarchy, and degree correlations.
\newblock \emph{Phys. Rev. E} 67:056104.

\bibitem{krapivsky_organization_2001}
Krapivsky PL, Redner S
\newblock (2001) Organization of growing random networks.
\newblock \emph{Phys. Rev. E} 63:066123.

\bibitem{krapivsky_network_2005}
Krapivsky PL, Redner S
\newblock (2005) Network growth by copying.
\newblock \emph{Phys. Rev. E} 71:036118.

\bibitem{gabel_highly_2013}
Gabel A, Krapivsky PL, Redner S
\newblock (2013) Highly dispersed networks by enhanced redirection.
\newblock \emph{Phys. Rev. E} 88:050802.

\bibitem{gabel_highly_2014}
Gabel A, Krapivsky PL, Redner S
\newblock (2014) Highly dispersed networks generated by enhanced redirection.
\newblock \emph{Journal of Statistical Mechanics: Theory and Experiment}
  2014:P04009.

\bibitem{WorldBankData}
(2015) {World Bank} data sources. (\url{http://data.worldbank.org/indicator})
\newblock Accessed: 2015-08.

\bibitem{petersen_quantitative_2014}
Petersen AM, Pavlidis I, Semendeferi I
\newblock (2014) A quantitative perspective on ethics in large team science.
\newblock \emph{Sci. \& Eng. Ethics.} 20:923--945.

\bibitem{Zhou200683}
Zhou P, Leydesdorff L
\newblock (2006) The emergence of china as a leading nation in science.
\newblock \emph{Research Policy} 35:83 -- 104.

\bibitem{PageRank:1998}
Page L, Brin S, Motwani R, Winograd T
\newblock (1998) {The PageRank Citation Ranking: Bringing Order to the Web}.,
  (Standford University), Technical report.

\bibitem{Dixon_Gini_1987}
Dixon PM, Weiner J, Mitchell-Olds T, Woodley R
\newblock (1987) Bootstrapping the {G}ini coefficient of inequality.
\newblock \emph{Ecology} 68:1548--1551.

\bibitem{Buldyrev_growth_2007}
Buldyrev SV, Growiec J, Pammolli F, Riccaboni M, Stanley HE
\newblock (2007) The growth of business firms: Facts and theory.
\newblock \emph{J. Eur. Economic Association} 5:574--584.

\bibitem{petersen_quantitative_2011}
Petersen AM, Jung WS, Yang JS, Stanley HE
\newblock (2011) Quantitative and empirical demonstration of the matthew effect
  in a study of career longevity.
\newblock \emph{Proceedings of the National Academy of Sciences} 108:18--23.

\bibitem{petersen_persistence_2012}
Petersen AM, Riccaboni M, Stanley HE, Pammolli F
\newblock (2012) Persistence and uncertainty in the academic career.
\newblock \emph{Proc. Natl. Acad. Sci. {USA}} 109:5213 -- 5218.

\bibitem{Scharnhorst_models_2012}
Scharnhorst A, B{\"o}rner K, van~den Besselaar P, {(Eds.)}
\newblock (2012) \emph{Models of Science Dynamics}
\newblock (Springer-Verlag Berlin Heidelberg, Berlin).

\bibitem{Golosovsky_transition_2013}
Golosovsky M, Solomon S
\newblock (2013) The transition towards immortality: Non-linear autocatalytic
  growth of citations to scientific papers.
\newblock \emph{Journal of Statistical Physics} 151:340--354.

\bibitem{Golosovsky_uncovering_2014}
Golosovsky M, Solomon S
\newblock (2014) Uncovering the dynamics of citations of scientific papers.
\newblock \emph{arXiv:1410.0343v1}.

\bibitem{PhysRevE.80.037101}
Wu ZX, Holme P
\newblock (2009) Modeling scientific-citation patterns and other triangle-rich
  acyclic networks.
\newblock \emph{Phys. Rev. E} 80:037101.

\bibitem{ren_modeling_2012}
Ren FX, Shen HW, Cheng XQ
\newblock (2012) Modeling the clustering in citation networks.
\newblock \emph{Physica A} 391:3533 -- 3539.

\bibitem{peterson_nonuniversal_2010}
Peterson GJ, Presse S, Dill KA
\newblock (2010) Nonuniversal power law scaling in the probability distribution
  of scientific citations.
\newblock \emph{Proc. Natl. Acad. Sci. {USA}} 107:16023--16027.

\bibitem{Golosovsky_stochastic_2012}
Golosovsky M, Solomon S
\newblock (2012) Stochastic dynamical model of a growing citation network based
  on a self-exciting point process.
\newblock \emph{Phys. Rev. Lett.} 109:098701.

\bibitem{wang_quantifying_2013}
Wang D, Song C, Barabasi AL
\newblock (2013) Quantifying long-term scientific impact.
\newblock \emph{Science} 342:127--132.

\bibitem{Simon_class_1955}
Simon HA
\newblock (1955) On a class of skew distribution functions.
\newblock \emph{Biometrika} 42:425--440.

\bibitem{Barabasi_evolution_2002}
Barabasi AL, {et~al.}
\newblock (2002) Evolution of the social network of scientific collaborations.
\newblock \emph{Physica A} 311:590 -- 614.

\bibitem{Jeong_Measuring_2003}
Jeong H, Neda Z, Barabasi AL
\newblock (2003) Measuring preferential attachment in evolving networks.
\newblock \emph{EPL} 61:567.

\bibitem{Redner2005PA}
Redner S
\newblock (2005) Citation statistics from 110 years of physical review.
\newblock \emph{Physics Today} 58:49--54.

\bibitem{Medo_statistical_2014}
Medo M
\newblock (2014) Statistical validation of high-dimensional models of growing
  networks.
\newblock \emph{Phys. Rev. E} 89:032801.

\bibitem{medo_temporal_2011}
Medo M, Cimini G, Gualdi S
\newblock (2011) Temporal effects in the growth of networks.
\newblock \emph{Phys. Rev. Lett.} 107:238701.

\bibitem{UnivCite}
Radicchi F, Fortunato S, Castellano C
\newblock (2008) Universality of citation distributions: Toward an objective
  measure of scientific impact.
\newblock \emph{Proc. Natl. Acad. Sci. USA} 105:17268--17272.

\bibitem{hirsch_index_2005}
Hirsch J
\newblock (2005) An index to quantify an individual's scientific research
  output.
\newblock \emph{Proc. Natl. Acad. Sci. {USA}} 102:16569 -- 16572.

\bibitem{petersen_statistical_2011}
Petersen AM, Stanley HE, Succi S
\newblock (2011) Statistical regularities in the rank-citation profile of
  scientists.
\newblock \emph{Scientific Reports} 1:181.

\bibitem{petersen_z-index:_2013}
Petersen AM, Succi S
\newblock (2013) The {Z}-index: A geometric representation of productivity and
  impact which accounts for information in the entire rank-citation profile.
\newblock \emph{Journal of Informetrics} 7:823 -- 832.

\bibitem{petersen_methods_2011}
Petersen AM, Penner O, Stanley HE
\newblock (2011) Methods for detrending success metrics to account for
  inflationary and deflationary factors.
\newblock \emph{Eur. Phys. J. B} 79:67--78.

\bibitem{simonton_after_2013}
Simonton DK
\newblock (2013) After {E}instein: Scientific genius is extinct.
\newblock \emph{Nature} 493:602--602.

\bibitem{petersen_quantifying_2015}
Petersen AM
\newblock (2015) Quantifying the impact of weak, strong, and super ties in
  scientific careers.
\newblock \emph{Proc. Natl. Acad. of Sci.} 112:E4671--E4680.

\bibitem{Wildson_need_2015}
Wildson J
\newblock (2015) We need a measured approach to metrics.
\newblock \emph{Nature} 523:129.

\bibitem{TheMetricTide}
Wilsdon J, {et~al.}
\newblock (2015) {The Metric Tide}: Report of the independent review of the
  role of metrics in research assessment and management., (Higher Education
  Funding Council for England {(HEFCE)}), Technical report.

\end{thebibliography}

\end{document}